\documentstyle[12pt]{article}
\input{epsf}
\newcommand{\be}{\begin{eqnarray}}
\newcommand{\ee}{\end{eqnarray}}
\newcommand{\ba}{\begin{array}}
\newcommand{\ea}{\end{array}}
\newcommand{\half}{{\textstyle{\frac{1}{2}}}}
\newcommand{\textfrac}[2]{{\textstyle{\frac{#1}{#2}}}}
\newcommand{\fourint}[1]{\int\!\frac{d^4 #1}{(2\pi)^4}}
\newcommand{\Fdual}{\widetilde{F}}
\newcommand{\cF}{{\cal F}}
\newcommand{\cR}{{\cal R}}

\newcommand{\cD}{{\cal D}}
\newcommand{\cU}{{\cal U}}
\newcommand{\tr}{{\rm tr}\,}

\newcommand{\nablaslash}{\nabla\hspace{-.65em}/\hspace{.3em}}
\newcommand{\Aslash}{A\hspace{-.65em}/\hspace{.3em}}
\newcommand{\Sslash}{S\hspace{-.65em}/\hspace{.3em}}
\newcommand{\partialslash}{\partial\hspace{-.5em}/\hspace{.15em}}
\newcommand{\kslash}{k\hspace{-.5em}/\hspace{.15em}}

\newcommand{\pslash}{p\hspace{-.5em}/\hspace{.15em}}
\newcommand{\llangle}{\left\langle}
\newcommand{\rrangle}{\right\rangle}
\newcommand{\intpsi}{\int {\cal D}\psi^\dagger {\cal D}\psi\,}

\newcommand{\effop}[1]{\mbox{``$#1$''}} 
\newcommand{\bftau}{\mbox{\boldmath{$\tau$}}}
\begin{document}
%
%
\rightline{RUB-TPII-6/97}
\rightline{July 1997}
\vspace{.3cm}
\begin{center}
\begin{large}
{\bf Nucleon matrix elements of higher--twist operators from the 
instanton vacuum} 
\\
\end{large}
\vspace{1.4cm}
{\bf J.\ Balla}$^{\rm a}$ , {\bf M.V.\ Polyakov}$^{\rm a, b}$ 
{\bf and C. Weiss}$^{\rm a}$  
\\[1.cm]
$^{a}$ {\em Institut f\"ur Theoretische Physik II,
Ruhr--Universit\"at Bochum, \\ D--44780 Bochum, Germany} \\
$^{b}$ {\em Petersburg Nuclear Physics Institute, Gatchina,
St.Petersburg 188350, Russia} 
\end{center}
\vspace{1.5cm}
\begin{abstract}
\noindent
We compute the nucleon matrix elements of QCD operators of twist 3 and 4 in
the instanton vacuum. We consider the operators determining $1/Q^2$--power
corrections to the Bjorken, Ellis--Jaffe and Gross--Llewellyn-Smith sum
rules.  The nucleon is described as a soliton of the effective chiral
theory derived from instantons in the $1/N_c$--expansion. QCD operators
involving the gluon field are systematically represented by effective
operators in the effective chiral theory. We find that twist--3 matrix
elements are suppressed relative to twist--4 by a power of the packing 
fraction of the instanton medium. Numerical results for the spin--dependent 
($d^{(2)}, f^{(2)}$) and spin--independent twist--3 and 4 
matrix elements are compared with results of other approaches and with 
experimental estimates of power corrections. The methods developed can 
be used to evaluate a wide range of matrix elements relevant to DIS.
\end{abstract}
\vspace{1cm}
\newpage 
\tableofcontents
\newpage
\section{Introduction}
\setcounter{equation}{0}
The discovery of scaling in deep--inelasting scattering experiments can be
seen as the starting point of modern strong interaction physics, giving
rise to the parton model and supporting the idea of asymptotic freedom.
QCD was established as a theory which describes logarithmic corrections to
scaling in the asymptotic region. The predictions of perturbative QCD for
the scale dependence of structure functions in leading and
next--to--leading order have been well confirmed by experiment However, in
general the $Q^2$--dependence of structure functions is subject to power
corrections, which lie outside the scope of perturbative QCD.  These
corrections are governed by scale parameters associated with 
non-perturbative vacuum fluctuations.
\par
The standard tool to analyze the scale dependence of structure functions in
QCD has been the operator product expansion. The part of nucleon structure
functions exhibiting only logarithmic scale dependence is determined by
nucleon matrix elements of operators of leading twist, while a class of
power corrections are associated with operators of non-leading twist
\cite{SV82}. In a partonic language twist--2 operators are one--particle
operators counting the contribution of individual partons to the structure
functions, while higher--twist operators can be interpreted as measuring
the effect of the chromoelectric and --magnetic field of the nucleon, or
effects of correlation between partons, on the structure functions. For a
quantitative description of power corrections one needs the values of the
matrix elements of these operators. These can be regarded as fundamental
characteristics of the nucleon. Clearly, an estimation of these quantities
requires a theory of the non-perturbative gluon degrees of freedom.
\par
A microscopic picture of the non-perturbative fluctuations of the gluon
field is provided by the instanton vacuum. The relevance of instantons to
the hadronic world was explored already before a quantitative theory of the
instanton medium existed \cite{tH76,CDG78,Sh82}.  A treatment of the
interacting instanton ensemble by means of the Feynman variational
principle showed how the instanton medium stabilizes itself
\cite{DP84}. The most prominent characteristic of the instanton vacuum is
the small packing fraction of the medium, {\em i.e.}, the small ratio of
the average size of the instantons in the medium to the average separation
between nearest neighbors. A value ${\bar\rho}/{\bar R} \simeq 1/3$ was
found both from phenomenological considerations \cite{Sh82} and from the
variational principle \cite{DP84}. This small parameter is the starting
point for a systematic analysis of non-perturbative phenomena in this
picture, such as chiral symmetry breaking, vacuum condensates {\em etc.}
We note that the instanton vacuum has also been studied in lattice
experiments using the cooling method. The measured properties of the
instanton medium found after cooling of the quantum fluctuations are close
to those computed from the variational principle \cite{PV88}.
\par
The main success of the instanton vacuum is its explanation of the
dynamical breaking of chiral symmetry, which is the most important
non-perturbative phenomenon determining the structure of light hadrons,
including the nucleon. The mechanism of chiral symmetry breaking is the
delocalization of the fermionic zero modes associated with the individual
instantons in the medium \cite{DP86,P89}, resulting in a finite fermion
spectral density at zero eigenvalue, which is proportional to the chiral
condensate \cite{BC80}. In an alternative formulation of chiral symmetry
breaking one derives the effective action of fermions in the instanton
medium by integrating over the instanton coordinates in the ensemble.  It
has the form of a Nambu--Jona-Lasinio model \cite{DP86_prep} with a
many--fermionic interaction with a specific spin--flavor structure, the
so-called 't Hooft determinant \cite{tH76}. Due to this interaction the
quarks develop a dynamical mass, and a Goldstone pion appears as a
collective excitation. The effective theory of massive quarks and pions
applies to quark momenta up the inverse instanton size, $\bar\rho^{-1}
\simeq 600 \, {\rm MeV}$, which acts as a natural cutoff.  The ratio of the
dynamically generated quark mass to the cutoff is directly related to the
packing fraction of the instanton medium, 
$M\bar\rho \propto (\bar\rho / R )^2 $. Thus, the diluteness of the
instanton medium ensures that the picture of interacting ``constituent''
quarks applies in a parametrically wide range\footnote{For a recent 
review of the instanton vacuum see \cite{SchSh96,D96_Varenna}.}.
\par
An immediate application of the effective chiral theory derived from
instantons is the quark--soliton model of baryons \cite{DPP88}.  In the
limit of a large number of colors, $N_c$, the nucleon can be viewed as
$N_c$ ``valence'' quarks bound by a self--consistent hedgehog--like pion
field. This approach gives a very reasonable description of a variety of
nucleon and $\Delta$ properties such as the $N$--$\Delta$ splitting,
electric formfactors, magnetic moments, axial coupling constants {\em
etc.}\ \cite{Review}.  Recently it was shown that this approach also allows
to compute the twist--2 quark distribution functions of the nucleon at a
low normalization point of order $\bar\rho^{-1}$
\cite{DPPPW96,DPPPW97,PP96}.
\par
The instanton vacuum, with the resulting effective chiral theory,
allows to evaluate hadronic matrix elements of QCD operators
involving the gluon field. In ref.\cite{DPW95} a method was developed
by which ``gluonic'' operators can systematically be represented as
effective many--fermionic operators in the effective chiral theory, the
hadronic matrix elements of which can then be computed using standard
techniques. This ``fermionization'' of QCD operators is possible relying 
entirely on the approximations which are already inherent in the 
effective theory, namely the diluteness of the instanton medium
and the $1/N_c$--expansion --- no additional assumptions are required.
It was shown in \cite{DPW95} that this approach preserves the essential
renormalization properties of QCD; for example, the QCD trace and $U(1)$ 
anomalies are realized in this approach at the level of hadronic matrix 
elements. This method is thus well suited for computing the matrix 
elements of the QCD operators which arise in the description of 
deep--inelastic scattering, both for leading and non-leading twist. 
\par
In this paper we evaluate the nucleon matrix elements of the operators of 
twist 3 and 4, which describe power corrections to the lowest moments
of polarized and unpolarized nucleon structure functions, using the 
approach of \cite{DPW95}. Our aims here are manifold. First,
we want to develop the methods for computing nucleon matrix of
operators of the type arising in the description of power corrections
to DIS, and demonstrate the consistency of this approach; for example, 
we verify that the QCD equations of motion are preserved by the 
effective operators. Second, we want to classify the various higher--twist 
matrix elements according to their magnitude with regard to the parameters 
of our approach, the instanton packing fraction, 
$M\bar\rho \propto (\bar\rho / R)^2$, and the formal parameter $1/N_c$. 
We shall see that a characteristic difference between operators
of highest twist (lowest spin) and lower twists (higher spins)
emerges as a consequence of the $O(4)$--symmetry of the instantons.
We also investigate the role of the non-perturbative gluon degrees
of freedom in twist--2 operators within our approach. 
Finally, we shall obtain numerical values for the higher--twist
matrix elements and compare with experiment as well as with other
theoretical estimates. The power corrections are sensitive to
the form of the non-perturbative vacuum fluctuations, so their study
and comparison with experiment will shed some light on the
srtructure of these fluctuations.
\par
The plan of this paper is as follows.
In section 2 we discuss the twist--3 and 4 operators whose matrix elements 
determine power corrections to the lowest moments of polarized and 
unpolarized nucleon structure functions. In section 3, we first summarize the 
properties of the instanton vacuum and the effective chiral theory.
We then derive the effective operators for the QCD operators
of leading and non-leading twist which generally arise in the description 
of nucleon structure functions. In the next step we compute the matrix 
elements of the twist--3 and 4 operators in quark states and discuss
the parametric suppression of twist 3 relative to twist 4 in the 
instanton packing fraction. Furthermore, we show explicitly that 
the QCD equations of motion are preserved by the effective operators. 
In section 4 we briefly describe the nucleon in the effective chiral 
theory. We then derive the expressions for the matrix elements 
of the gluonic operators of interest. We classify the spin--dependent and 
independent higher--twist matrix elements according to their magnitude 
in the $1/N_c$--expansion. We then compute the $N_c$--leading higher--twist
matrix elements and give crude estimates for the subleading ones.
A discussion of the numerical results and comparison with experimental
numbers and results of other non-perturbative methods are presented in 
section 5. Section 6 offers conclusions and an outlook.
\section{Power corrections to DIS}
\setcounter{equation}{0}
The $Q^2$--dependence of nucleon structure functions in QCD is usually
investigated using the operator product expansion \cite{SV82}. Expanding
the product of electromagnetic or weak currents in the forward Compton
amplitude at near light--cone separations in operators of increasing twist,
one can identify the scaling and power--suppressed parts of moments of the
structure functions.  The leading ($1/Q^2$--) power corrections to the
isovector and isosinglet combinations of the first moment of the polarized
structure function $g_1$ (the Bjorken \cite{Bj66} and Ellis--Jaffe
\cite{EJ74} sum rules) are given by\footnote{We note that there is a
mistake in ref.\cite{SV82} concerning the coefficients in this formula, see
\cite{JiU94,EhrMS94}.} \cite{EhrMS94}
\be
\int_0^1 dx \, g_1^{p - n} (x, Q^2 ) &=& \frac{1}{3} \left[
\frac{1}{2} a^{(0)}_{NS}
+ \frac{M_N^2}{9 Q^2} \left( a^{(2)}_{NS} + 4 d^{(2)}_{NS} + 4 f^{(2)}_{NS}
\right) + O\left(\frac{M_N^4}{Q^4}\right) \right] , 
\label{g1_NS} \\
\int_0^1 dx \, g_1^{p + n} (x, Q^2 ) &=& \frac{5}{9} \left[ 
\frac{1}{2} a^{(0)}_{S}
+ \frac{M_N^2}{9 Q^2} \left( a^{(2)}_S + 4 d^{(2)}_S + 4 f^{(2)}_S
\right) + O\left(\frac{M_N^4}{Q^4}\right) \right] .
\label{g1_S}
\ee
Here $M_N$ denotes the nucleon mass. The constants $a^{(0)}, a^{(2)}$ are
defined by the matrix elements of the twist--2 operators of spin 1 and 3,
respectively. More generally,
\be
\lefteqn{
\langle P S T_3 | \bar\psi \tau^3 \gamma_{\left\{\alpha \right.}
\gamma_5 \nabla_{\beta_1} \ldots \nabla_{\left. \beta_n \right\}} \psi  
| P S T_3 \rangle - \mbox{traces} } && \nonumber \\
&=& - 2 M_N (2 T_3 ) \; a^{(n)}_{NS} \;
S_{\left\{\alpha \right.} P_{\beta_1} \ldots P_{\left. \beta_n \right\}} 
- \mbox{traces} ,
\label{a2_NS} 
\\
\lefteqn{ \langle P S | \bar\psi \gamma_{\left\{\alpha \right.}
\gamma_5 \nabla_{\beta_1} \ldots \nabla_{\left. \beta_n \right\}} \psi  
| P S \rangle - \mbox{traces} } && \nonumber \\
&=& - 2 M_N \; a^{(n)}_{S} \;
S_{\left\{\alpha \right.} P_{\beta_1} \ldots P_{\left. \beta_n \right\}} 
- \mbox{traces} ,
\label{a2_S}
\\
&& (n = 0, 2, \ldots ) . \nonumber
\ee
The braces denote complete symmetrization in the respective indices. Here
$P$ is the nucleon 4--momentum, $S$ the nucleon polarization vector,
\be
S_\mu P^\mu &=& 0, \hspace{2em} S_\mu S^\mu \; = \; -1 .
\ee
The minus sign in eqs.(\ref{a2_NS}, \ref{a2_S}) relative to
ref.\cite{EhrMS94} comes because our $\gamma_5$ differs from the one of
Bjorken and Drell by a minus sign; this definition is used in
refs.\cite{DP86,DPP88,DPW95} and also ref.\cite{BK87}. In particular,
$a^{(0)}_{NS}$ and $a^{(0)}_{S}$ are identical to the nucleon isovector and
isosinglet axial coupling constants
\be
a^{(0)}_{NS} &=& g_A^{(3)}, \hspace{2cm} 
a^{(0)}_{S} \; = \; g_A^{(0)}.
\ee
\par
In eq.(\ref{a2_NS}) $T_3 = \pm 1/2$ for proton and neutron, respectively, and
$\tau^3$ denotes the Pauli matrix in flavor indices.
In the following it will always be understood that the flavor--nonsinglet
$(NS)$ and singlet $(S)$ parts of bilinear operators $\bar\psi \ldots \psi$
and their matrix elements are defined as in eqs.(\ref{a2_NS}, \ref{a2_S});
we shall not write the flavor dependence of matrix elements explicitly.
(We consider here the case of two light flavors; the generalization of the
formulas here and below to $SU(3)$--flavor is straightforward.)
\par
Of the power corrections in eqs.(\ref{g1_NS}, \ref{g1_S}), the terms
proportional to the twist--2 matrix element $a^{(2)}$ originate from the
expansion of Nachtmann moments (target mass corrections) \cite{SV82}, while
the other terms represent contributions of operators of non-leading
twist. The constant $d^{(2)}$ parametrizes the matrix element of the
twist--3 spin--2 operator \cite{EhrMS94}
\be
\lefteqn{ \langle P S | 
\bar\psi  \left( \gamma^\alpha \Fdual^{\beta\gamma}  
+ \gamma^\beta \Fdual^{\alpha\gamma}  \right) \psi  
| P S \rangle - \mbox{traces}} && \nonumber \\
&=& 2 M_N \; d^{(2)} \; \left[ 2 P^\alpha P^\beta S^\gamma
- P^\gamma P^\beta S^\alpha - P^\alpha P^\gamma S^\beta
+ (a \leftrightarrow \beta ) - \mbox{traces} \right] ,
\label{d2_def}
\ee
for both flavor non-singlet and singlet, see eqs.(\ref{a2_NS}, \ref{a2_S}).
It is understood that trace terms are to be subtracted on both sides, since
one is interested only in the twist--3 part.  The same operator contributes
also to the non-power suppressed--part of the third moment of the structure
function $g_2$,
\be
\int_0^1 dx \, x^2 g_2^{p - n} (x, Q^2 ) 
&=& \frac{1}{3} \left[ -\frac{1}{3} a^{(2)}_{NS}
+ \frac{1}{3} d^{(2)}_{NS} + O\left(\frac{M_N^2}{Q^2}\right) \right] , 
\label{g_2_NS} \\
\int_0^1 dx \, x^2 g_2^{p + n} (x, Q^2 ) 
&=& \frac{5}{9} \left[ -\frac{1}{3} a^{(2)}_S
+ \frac{1}{3} d^{(2)}_S + O\left(\frac{M_N^2}{Q^2}\right) \right] .
\label{g_2_S}
\ee
Finally, $f^{(2)}$ is given by the matrix element of the twist--4 spin--1
operators \cite{EhrMS94}
\be
\langle P S | \bar\psi \gamma_\alpha \Fdual^{\beta\alpha}
\psi | P S \rangle &=& 2 M_N^3 \; f^{(2)} \; S^\beta ,
\label{f2_def}
\ee
again for both flavor non-singlet and singlet. 
\par
In unpolarized neutrino scattering there are $1/Q^2$--power corrections to
the first moment of the isovector parity--even structure function $F_1$ and
the isosinglet parity--odd structure function $F_3$
(Gross--Llewellyn--Smith sum rule \cite{GLS69}),
\be
\frac{1}{2} \int_0^1 dx \, \left( 1 - \frac{2 M_N^2}{3 Q^2} x^2 \right)
F_{1, \nu}^{p - n} (x, Q^2 ) &=& -\frac{1}{2} 
\left[ 1 - \frac{8 M_N^2}{9 Q^2} c^{(2)}_{NS} 
+ O\left(\frac{M_N^4}{Q^4}\right)\right] ,
\label{F1_NS}
\\
\frac{1}{2} 
\int_0^1 dx \, \left( 1 - \frac{2 M_N^2}{3 Q^2} x^2 \right)
F_{3, \nu}^{p + n} (x, Q^2 ) &=& 3 \left[ 1  
- \frac{8 M_N^2}{27 Q^2} c^{(2)}_{S} + O\left(\frac{M_N^4}{Q^4}\right)
\right] .
\label{F3_S}
\ee
Here, $c^{(2)}$ is determined by the spin--independent nucleon matrix
element of the twist--4 spin--1 operator\footnote{We follow here
ref.\cite{BK87}, with 
$\langle\langle O^{NS, S} \rangle\rangle = M_N^2 c^{(2)}_{S, NS}$. Note
that the conventions of \cite{SV82} differ from those of \cite{BK87}; see
the corresponding remarks in ref.\cite{BK87}.}  \cite{BK87},
\be
\half \sum_{\rm spin}
\langle P | \bar\psi \gamma_\alpha \gamma_5 \Fdual^{\beta\alpha} 
\psi | P \rangle &=& 2 M_N^2 \; c^{(2)} \; P^\beta .
\label{c2_def}
\ee
\par
We note that in both the polarized and unpolarized case, eqs.(\ref{g1_NS},
\ref{g1_S}) and eqs.(\ref{F1_NS}, \ref{F3_S}), there are, of course,
logarithmic corrections, which we have not written down \cite{SV82}. They
originate from the scale dependence of the coefficients of the OPE, 
which is described by the renormalization group equation.
\par
In this paper we evaluate the nucleon matrix elements of the twist--3 and 
4 operators, eqs.(\ref{d2_def}, \ref{f2_def}) and eq.(\ref{c2_def}),
in the instanton vacuum. The matrix elements depend on the normalization
point of the operators, which in our approach is of the order of the
inverse average instanton size, $\bar\rho^{-1} \simeq 600 \, {\rm MeV}$.
Values at other normalization points can easily be obtained using 
the renormalization group equation.
\par
We emphasize that we are not dealing here with the instanton--induced
contributions to DIS which arise from non-perturbative power corrections
to the coefficient functions of the OPE.
Those are the leading semiclassical corrections associated with 
small instantons of size $\rho \sim 1/Q$ \cite{BB93,Moch96}.
\section{From QCD to the effective chiral theory}
\setcounter{equation}{0}
\subsection{Effective chiral theory from the instanton vacuum}
The basis of our description is the medium of independent instantons with
an effective size distribution, which was obtained by Diakonov and Petrov
as a variational approximation to the interacting instanton partition
function \cite{DP84}. In the large--$N_c$ limit the width of the effective
size distribution is of order $1/b \sim 1/N_c$, so we may assume instantons
and antiinstantons of fixed size,
\be
\bar\rho &\simeq& (600 \, {\rm MeV})^{-1} .
\ee
It is sufficient to consider a medium with equal number of instantons and
antiinstantons, $N_+ = N_- = N/2$, since we shall not be concerned with
topological fluctuations or effects related to the $U(1)$--anomaly here.
\par
Fermions couple to the instantons and antiinstantons ($I$'s and $\bar I$'s
in the following) mainly through the zero modes. In the zero--mode
approximation, which is justified by the diluteness of the instanton
medium, the interaction of the fermion field with given flavor $f$ with one
$I$ ($\bar I$), with collective coordinates $z$ (center) and $\cU$ (color
orientation) is given by\footnote{To avoid confusion with the pion field we
denote in this paper the color orientation matrices of the instanton in the
fundamental representation of $SU(N_c )$ by $\cU$, not by $U$ as in
ref.\cite{DPW95}.} \cite{DP86,DPW95}
\be
V_\pm [\psi_f^\dagger , \psi_f ] 
&=& 4\pi^2\bar\rho^2
\fourint{k_1}\fourint{k_2} e^{i (k_2 - k_1 ) \cdot z} \;
F(k_1 ) F(k_2 ) \nonumber \\
&& \times \; \psi_f^\dagger (k_1 ) \;
\left[ \frac{1}{8}
\gamma_\kappa \gamma_\lambda \frac{1 \pm \gamma_5}{2} \right] \;
\left[ \cU \tau^\mp_\kappa \tau^\pm_\lambda \cU^\dagger 
\right] \;
\psi_f (k_2 ) .
\label{V_I}
\ee
(We use the standard $\psi^\dagger = i\bar\psi$ for the quark field.)  Here
$F(k)$ is a form factor proportional to the Fourier transform of the wave
function of the fermion zero mode, which drops to zero for
momenta of order $\bar\rho^{-1}$,
\be
F(k) &=& - t \frac{d}{dt} \left[ I_0 (t) K_0 (t) - I_1 (t) K_1 (t)
\right] \; \rightarrow
\left\{ \ba{ll} 1 & t \rightarrow 0 \\  \textfrac{3}{4} t^{-3}
& t \rightarrow \infty \ea \right. 
\label{F_k} \\
t &=& \half k \bar\rho . \nonumber
\ee
The matrices $\tau^\pm_\kappa$ are 
$N_c \times N_c$ matrices with $(\bftau, \mp i)$ in the upper left corner 
and zero elsewhere.
\par
The effective fermion action ($N_f$ flavors), which is derived by
integrating over the coordinates of the instantons in the medium and
finding the saddle point of the fermion integral in the large--$N_c$ limit
\cite{DP86_prep,DPW95}, is of the form
\be
S_{\rm eff} [\psi^\dagger , \psi ] &=& - \left( \int d^4 x
\sum_f^{N_f} \psi^\dagger_f i\partialslash \psi_f
\; + \; Y_+ \; + \; Y_- \right) .
\label{S_eff} 
\ee
Here, $Y_\pm$ denote the $2 N_f$--fermion vertices 
\be
Y_\pm [\psi^\dagger , \psi ]
&=& \left( i \frac{N_c M}{4\pi^2\bar\rho^2} \right)^{N_f}
\int d^4 z \int d\cU \; \prod_{f = 1}^{N_f} 
V_\pm [\psi_f^\dagger , \psi_f ] 
\nonumber \\
&=& \left(\frac{2V}{N}\right)^{N_f - 1} (i M)^{N_f} 
\int d^4 x \, \det J_\pm (x) ,
\label{Y}
\ee
where $J_\pm (z)_{fg}$ are color singlet currents which are 
$N_f \times N_f$--matrices in flavor,
\be
J_\pm (x)_{fg} &=& \fourint{k}\fourint{l} e^{-i(k - l)\cdot x}
\, F(k) F(l) \; \psi^\dagger_f (k) \frac{1 \pm \gamma_5}{2} \psi_g (l) ,
\label{J}
\ee
and the determinant in eq.(\ref{Y}) is over flavor indices.  The vertices
eq.(\ref{Y}) have the spin--flavor structure of the 't Hooft determinant
\cite{tH76}. The dynamical quark mass, $M$, is determined by the
saddle--point equation.  Its square is proportional to the instanton
density, since chiral symmetry breaking is a collective effect involving
all instantons in the medium. Thus, parametrically,
\be
(M\bar\rho )^2 &\propto& \frac{N}{V N_c} \bar\rho^4 
\;\; \propto \;\; \left(\frac{\bar\rho}{R}\right)^4 .
\label{M_rho_to_rho_R}
\ee
The diluteness of the instanton medium thus ensures that the dynamical
quark mass is small compared to the momentum cutoff, $\bar\rho^{-1}$, so
that the effective theory of massive quarks described by eq.(\ref{S_eff})
is valid in a parametrically wide range.  This fact is central to the
concept of the effective chiral action. 
\par
In the case of one quark flavor, $N_f = 1$, the 't Hooft
vertex, eq.(\ref{Y}), is simply a mass term, and the effective
fermion action becomes
\be
S_{\rm eff} [\psi^\dagger , \psi ]_{N_f = 1} &=& \fourint{k} 
\psi^\dagger (k)
\left[ \kslash - i M F^2 (k) \right] \psi (k) .
\label{S_eff_1f}
\ee
For more than one quark flavors, $N_f > 1$, the effective action,
eq.(\ref{S_eff}), describes a system of quarks with many--fermionic
interactions. The partition function of this theory can be written in
bosonized form by introducing meson fields \cite{DP86_prep,DPW95}.
\be
Z &=& \int \cD \pi \intpsi
\exp \int d^4 x \left( \sum_f^{N_f} \psi^\dagger_f
i \partialslash \psi_f \right. \nonumber \\
&& \left. + i M \sum_{f, g}^{N_f} 
\left[ U_+ (x)_{fg} J_+ (x)_{fg} + U_- (x)_{fg} J_- (x)_{fg} \right] 
\right) . 
\label{Z_bosonized}
\ee
Here, $\pi (x)$ is the pion field, and
\be
U_+ (x) &=& U(x), \hspace{2cm} U_- (x) \;\; = \;\; U^\dagger (x), 
\nonumber \\
U (x) &=& \exp\left[ i\pi^a (x) \tau^a \right] .
\label{U}
\ee
We have retained in eq.(\ref{Z_bosonized}) only the Goldstone degrees of
freedom; other mesons ({\em e.g.}\ rho, sigma) have masses of order of the
cutoff, $\bar\rho^{-1}$, and thus should not be considered as dynamical
degrees of freedom in this effective theory. Eq.(\ref{Z_bosonized})
describes the minimal chirally invariant coupling of quarks to Goldstone
bosons.
\par
Eq.(\ref{Z_bosonized}) is the starting point for the description of the 
nucleon, which in the large--$N_c$ limit is described by a classical pion 
field of hedgehog form, see Section \ref{section_nucleon}.
\subsection{Effective gluon operators}
\label{subsec_effective_operators}
The higher--twist operators arising in the description of power corrections 
to DIS explicitly involve the gluon field. Following the approach of
\cite{DPW95} we now determine the effective fermion operators representing 
these operators in the effective chiral theory. 
\par
In a dilute medium of instantons the interaction of a gluonic operator
with the light fermions is dominated by the contribution of a single 
$I (\bar I)$; many--instanton contributions are suppressed by powers 
of the packing fraction. Thus, when constructing the effective operator
corresponding to a given function of the gauge field, one evaluates
this function in the field of one $I (\bar I)$. The functions of the 
gauge field appearing in the QCD operators of twist 2, 
eqs.(\ref{a2_NS}, \ref{a2_S}), and twist 3 and 4, eqs.(\ref{d2_def}, 
\ref{f2_def}, \ref{c2_def}), have the property that in the field of one 
$I (\bar I)$ they are proportional to the 't Hooft eta symbol. 
For instance, the field strength and dual field strength of one $I$ 
($\bar I$) in singular gauge have the form
\be
F^a_{\beta\gamma} (x)_{I (\bar I)} &=& (\eta^\mp )^a_{\mu\nu} 
\frac{8  \rho^2 }{(x^2 + \rho^2 )^2}
\left( \frac{x_\mu x_\beta}{x^2} \delta_{\gamma\nu} 
+ \frac{x_\nu x_\gamma}{x^2} \delta_{\mu\beta} 
- \half \delta_{\mu\beta} \delta_{\gamma\nu} \right), 
\label{F_inst} \\
\Fdual^a_{\beta\gamma} (x)_{I (\bar I)} &=& 
\pm F^a_{\beta\gamma} (x)_{I (\bar I)} ,
\label{Fdual_inst}
\ee
where $(\eta^-)^a_{\mu\nu} = \bar\eta^a_{\mu\nu},  
(\eta^+)^a_{\mu\nu} = \eta^a_{\mu\nu}$.
The same applies to the gauge field itself, which is given by
\be
A^a_\beta (x)_{I (\bar I)} &=& (\eta^\mp )^a_{\mu\nu} 
\frac{2 \rho^2}{(x^2 + \rho^2 ) x^2} x_\nu \delta_{\mu\beta} ,
\label{A_inst}
\ee
or, more generally, to the products of gauge fields which arise
when multiplying out the product of covariant derivatives in the
twist--2 operators of arbitrary spin, eqs.(\ref{a2_NS}, \ref{a2_S}),
\be
\left( A^{a_1}_{\beta_1} (x) \ldots A^{a_n}_{\beta_n} (x) 
\right)_{I (\bar I)}
\tr \left[ \lambda^a \frac{\lambda^{a_1}}{2} \ldots 
\frac{\lambda^{a_n}}{2} \right]
&=& (\eta^\mp )^a_{\mu\nu} 
f_{\pm , \mu\nu \beta_1 \ldots \beta_n} (x) ,
\ee
where $f_{\pm \mu\nu \beta_1 \ldots \beta_n}(x)$ denotes a Lorentz tensor
depending on $x$ whose precise form can be determined by explicitly 
computing the product using eq.(\ref{A_inst}).
\par
To account for all cases of interest\footnote{We note that there 
are functions of the gauge field in the adjoint representation which 
in one $I (\bar I)$ are 
not proportional to the eta symbol. They are constructed
with the help of the totally symmetric structure constants, $d^{abc}$.
These functions lead to effective operators of a form different from 
the one described below. We do not need to consider them here.}, 
we consider in the following a general ``gluonic'' operator of the form
\be
O_{\alpha_1 \ldots \alpha_r \beta_1 \ldots \beta_s}
 (x) &=& \bar\psi (x) \frac{\lambda^a}{2} 
\Gamma_{\alpha_1 \ldots \alpha_r} \, \psi (x) \;
\cF^a_{\beta_1 \ldots \beta_s} (x) ,
\label{O}
\ee
where $\Gamma_{\alpha_1 \ldots \alpha_r}$ is a matrix in Dirac spinor 
indices and $\cF^a_{\beta_1 \ldots \beta_s} (x) \equiv 
\cF^a_{\beta_1 \ldots \beta_r} [ A(x) ]$ denotes a function of the gauge 
field in the adjoint representation, which in the field of one 
$I (\bar I)$, in standard orientation and centered at zero, is given by
\be
\cF^a_{\beta_1 \ldots \beta_r} (x)_{I (\bar I)} &=& (\eta^\mp )^a_{\mu\nu} 
\; \cF_{\pm , \mu\nu \beta_1 \ldots \beta_r} (x), 
\label{F_adjoint}
\ee
with $\cF_{\pm , \mu\nu \beta_1 \ldots \beta_r} (x)$ are general 
tensor functions of the coordinate, $x$. For a general instanton 
with center $z$ and color orientation given by the 
$SU(N_c)$ matrix $\cU$, {\em cf.}\ eq.(\ref{V_I}), the function 
eq.(\ref{F_adjoint}) takes the form
\be
\cF^a_{\beta_1 \ldots \beta_r} (x)_{I (\bar I)} &=& 
\half \tr [\lambda^a \cU \lambda^b \cU^\dagger ] 
(\eta^\mp )^b_{\mu\nu} \;
\cF_{\pm , \mu\nu \beta_1 \ldots \beta_r} (x - z).
\label{F_adjoint_U_z}
\ee
Following \cite{DPW95} we define a $2 N_f$--fermion vertex as the average 
of the product of the function eq.(\ref{F_adjoint_U_z}) and the 
instanton--fermion interaction potential, eq.(\ref{V_I}), over the 
collective coordinates of one instanton,
\be
\lefteqn{ 
(Y_{\cF\pm})^a_{\beta_1 \ldots \beta_r} [\psi^\dagger , \psi ] (x) } && 
\nonumber \\ &=& \left(\frac{2V}{N}\right)^{N_f - 1}
\left( i \frac{N_c M}{4\pi^2\bar\rho^2} \right)^{N_f} \int d^4 z 
\int d\cU \; 
\cF^a_{\beta_1 \ldots \beta_r} (x)_{I (\bar I)} \; \prod_{f = 1}^{N_f} 
V_\pm [\psi_f^\dagger , \psi_f ] ,
\label{Y_F}
\ee
Performing the integral over color orientations in the leading order of 
$1/N_c$, as in eq.(\ref{Y}), one obtains
\be
\lefteqn{
(Y_{\cF\pm})^a_{\beta_1 \ldots \beta_r} [\psi^\dagger , \psi ] (x) } 
&& \nonumber \\
&=& \left(\frac{2V}{N}\right)^{N_f - 1} \frac{(i M)^{N_f}}{N_c}
\int d^4 z \, \cF_{\pm , \mu\nu \beta_1 \ldots \beta_r} (x - z) \, 
\sum_{f, g = 1}^{N_f} J^a_{\pm , \mu\nu} (z)_{fg} \;
{\textstyle{\det'_{fg}}} J_\pm (z) .
\label{Y_F_J}
\ee
Here $J_\pm (z)$ are the left-- and right-handed color singlet
currents, eq.(\ref{J}), while $J^a_{\pm , \mu\nu}$ is the color--octet 
Lorentz tensor current,
\be
J^a_{\pm\mu\nu} (x)_{fg} &=& 
\fourint{k}\fourint{l} e^{-i(k - l) \cdot x} 
\, F(k) F(l) \; \psi^\dagger_f (k) \frac{\lambda^a}{2}
\sigma_{\mu\nu}
\frac{1 \pm \gamma_5}{2} \psi_g (l) ,
\nonumber \\
\label{J_sigma}
\ee
where $\sigma_{\mu\nu} = i \half [\gamma_\mu , \gamma_\nu ]$.  In
eq.(\ref{Y_F_J}) $\det'_{fg}$ denotes the minor of the determinant in which
the $f$--th row and $g$--th column are omitted.  The form of the vertex
eq.(\ref{Y_F_J}) is intuitively plausible: the ``effective gluonic field''
is given by the color--octet current of the quark fields convoluted with
the function of the field of one instanton. Note that the vertex
eq.(\ref{Y_F_J}) is chirally invariant --- the tensor currents,
eq.(\ref{J_sigma}), have the same chirality properties as the
scalar--pseudoscalar ones, eq.(\ref{J}), and with respect to flavor the
terms in eq.(\ref{Y_F_J}) combine to give the determinant, which is an
invariant.
\par
The effective operator which represents eq.(\ref{O}) in the effective
chiral theory is given by \cite{DPW95,PW95}
\be
\lefteqn{
\effop{O}_{\alpha_1 \ldots \alpha_r \beta_1 \ldots \beta_s}
[\psi^\dagger , \psi ] (x) } && \nonumber \\
&=& 
-i \psi^\dagger (x) \frac{\lambda^a}{2} \Gamma_{\alpha_1 \ldots \alpha_r} 
\psi (x) \left[
(Y_{\cF +})^a_{\beta_1 \ldots \beta_s} (x) N_+ \cR_+ 
+ (Y_{\cF -})^a_{\beta_1 \ldots \beta_s} (x) N_- \cR_- \right] .
\label{effop}
\ee
Here $N_+ = N_- = N/2$ are the numbers of $I's$ and $\bar I's$ in the
ensemble. The $\cR_{\pm}$ are many--fermionic
vertices which act as the ``inverse'' of the 't Hooft vertex, eq.(\ref{Y}),
in saddle--point approximation (large--$N_c$ limit); symbolically
\be
\cR_\pm &=& Y_\pm^{-1} .
\label{symbolic}
\ee
(An explicit representation for $\cR_\pm [\psi^\dagger , \psi ]$ 
can be found in \cite{DPW95}.)
Eq.(\ref{symbolic}) is to be understood in the following sense. 
The vacuum expectation value of the vertices is
\be
\llangle \cR_\pm \rrangle &=& \llangle Y_\pm \rrangle^{-1} 
\;\; = \;\; N_\pm^{-1} ,
\label{R_vev}
\ee
while the connected average of $\cR_\pm$ with other fermion fields is
equivalent to the insertion of a 't Hooft vertex, eq.(\ref{Y}),
\be
\llangle \psi (y') \ldots \cR_\pm \; \psi^\dagger (y) \ldots 
\rrangle_{\rm conn} 
&=& - \llangle Y_\pm \rrangle^{-2} 
\llangle \psi (y') \ldots Y_\pm \; \psi^\dagger (y) \ldots 
\rrangle_{\rm conn} \nonumber \\
&=& - N_\pm^{-2} 
\llangle \psi (y') \ldots Y_\pm \; \psi^\dagger (y) \ldots 
\rrangle_{\rm conn} .
\label{R_conn}
\ee
We note that the operator character of the vertices $\cR_\pm$,
eq.(\ref{symbolic}), is relevant only when computing connected averages of
fermion fields with effective operators which have a non-zero vacuum
expectation value, such as the operator $\bar\psi \Aslash \psi$, see
section \ref{subsec_eom}. For operators with zero vacuum expectation value,
which we shall mostly be concerned with in the following, the vertices
$\cR_{\pm}$ in eq.(\ref{effop}) can be replaced by their vacuum expectation
values, eq.(\ref{R_vev}), so that the effective operator eq.(\ref{effop})
reduces to
\be
\lefteqn{\effop{O}_{\alpha_1 \ldots \alpha_r \beta_1 \ldots \beta_s} 
[\psi^\dagger , \psi ] (x) } && \nonumber \\
&=& -i \psi^\dagger (x) \frac{\lambda^a}{2} 
\Gamma_{\alpha_1 \ldots \alpha_r} \psi (x) 
\left[ (Y_{\cF +})^a_{\beta_1 \ldots \beta_s} (x) 
+ (Y_{\cF -})^a_{\beta_1 \ldots \beta_s} (x) \right] .
\label{effop_novev}
\ee
(Diagrams in which the factors $\cR_\pm$ are connected to external quark
lines are suppressed by $1/N_c$ relative to the disconnected terms.)
\par
The effective operator eq.(\ref{effop_novev}) is given by the sum of
$I$ and $\bar I$ contributions. If the function of the gluon field, 
eq.(\ref{F_adjoint}), takes the same value for $I$ and $\bar I$, 
$\cF_{-, \mu\nu\beta_1 \ldots \beta_s} = 
\cF_{+, \mu\nu\beta_1 \ldots \beta_s}$, as for example in the case
of the field strength, $F_{\mu\nu}$, eq.(\ref{F_inst}) or the gauge 
field, $A_\mu$, eq.(\ref{A_inst}), the 
sum of vertices $(Y_{\cF \pm})^a_{\beta_1 \ldots \beta_s}$ is an 
operator of natural parity (true tensor). If 
$\cF_{\pm , \mu\nu\beta_1 \ldots \beta_s}$ takes opposite values 
for $I$'s and $\bar I$'s, as for example for the dual field, 
$\Fdual_{\mu\nu}$, 
eq.(\ref{Fdual_inst}), the sum of vertices represents an operator 
of unnatural parity (pseudotensor).
\par
\begin{figure}
\vspace{-1cm}
\epsfxsize=15cm
\epsfysize=12cm
\epsffile{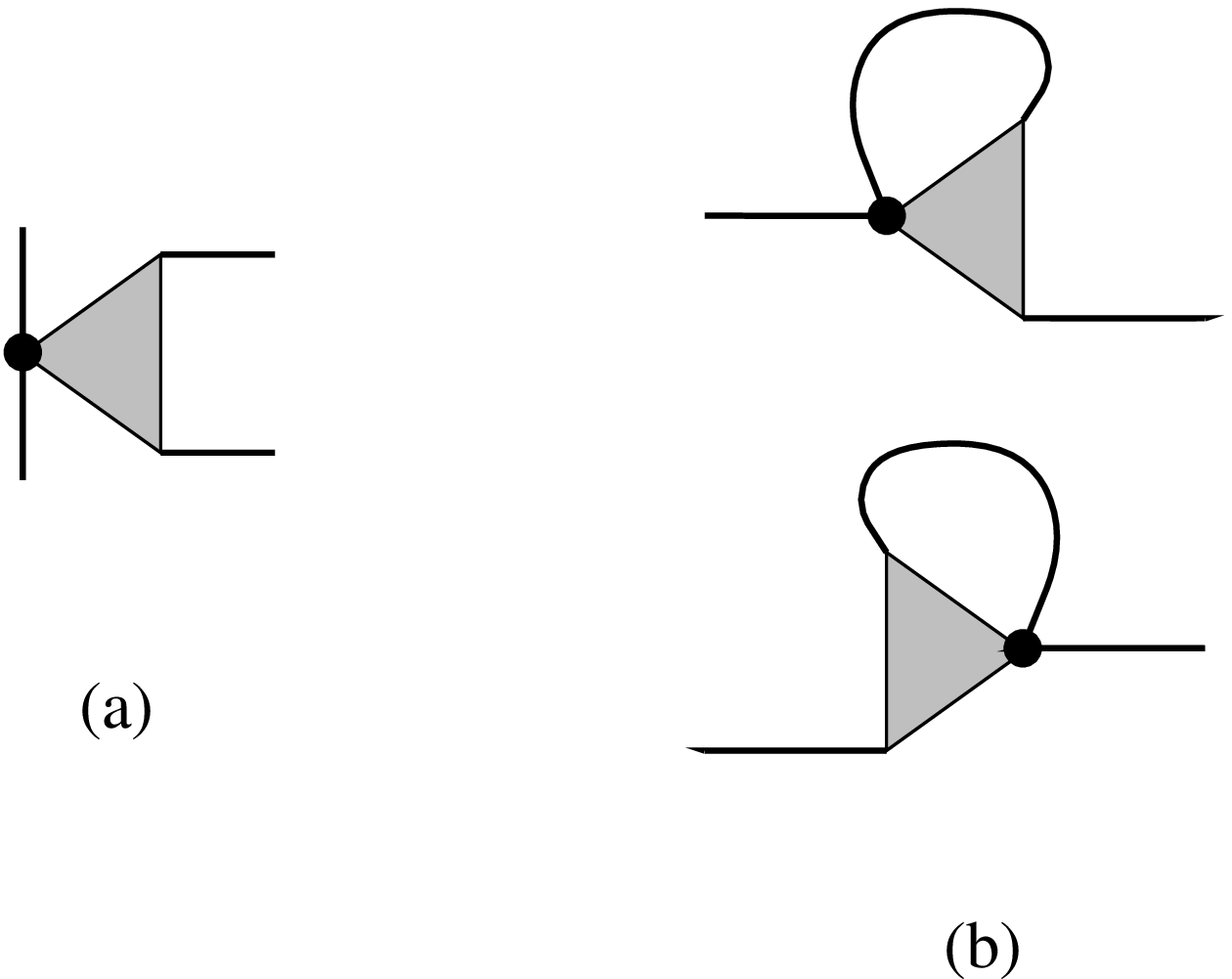}
\caption[]{(a) The effective operator, eq.(\ref{effop_novev}), in the case
of one quark flavor, $N_f = 1$. (b) Diagrams contributing to the matrix 
element of the effective operator in quark states.}
\label{fig_quark}
\end{figure}

In the case of one quark flavor, $N_f = 1$, the fermion vertex, 
eq.(\ref{Y_F}), is bilinear in the fermion field,
\be
(Y_{\cF\pm})^a_{\beta_1 \ldots \beta_s} [\psi^\dagger , \psi ] 
(x)_{N_f = 1} 
&=& \frac{i M}{N_c} \int d^4 z \, 
\cF_{\pm , \mu\nu \beta_1 \ldots \beta_s} (x - z) 
J^a_{\pm\mu\nu} (z) .
\label{Y_F_Nf1}
\ee
where $J_{\pm \mu\nu}$ are the left-- and right--handed currents, 
eq.(\ref{J_sigma}). The vertex can be expressed in momentum representation as
\be
\lefteqn{
(Y_{\cF\pm})^a_{\beta_1 \ldots \beta_s} [\psi^\dagger , \psi ] 
(x)_{N_f = 1} } \nonumber \\
&=& \frac{i M}{N_c} \fourint{k_1} \fourint{k_2}
e^{i (k_2 - k_1 ) \cdot x}
F(k_1 ) F(k_2 ) \; \cF_{\pm , \mu\nu \beta_1 \ldots \beta_s} (k_2 - k_1 ) 
\nonumber \\
&& \times \; \psi^\dagger (k_1 ) \frac{\lambda^a}{2}
\left[ \sigma_{\mu\nu} \frac{1 \pm \gamma_5}{2} \right]
\psi (k_2 ) .
\label{Y_F_mom}
\ee
Here the Fourier transform of the function of the instanton field,
eq.(\ref{F_adjoint}), is defined as
\be
\cF_{\pm, \mu\nu \beta_1 \ldots \beta_s} (k)
&=& \int d^4 x \, e^{-ik\cdot x} \;
\cF_{\pm , \mu\nu \beta_1 \ldots \beta_s} (x) .
\label{F_Fourier}
\ee
The effective operator, eq.(\ref{effop_novev}), in this case is a
four--fermionic operator. It has the form of a color--octet
current--current interaction, with one current formed by the fermion fields
present in the original QCD operator, the other by the non-local
instanton--induced vertices. Graphically it can be represented as in
Fig.(\ref{fig_quark}).
\par
In the case of more than one quark flavor, $N_f > 1$, the effective
operator is obtained at first as a $(2 N_f + 2)$--fermionic operator,
eq.(\ref{effop_novev}). When passing to the bosonized effective theory,
eq.(\ref{Z_bosonized}), one may convert also the effective operator to a
simple ``bosonized'' form. Consider a correlation function with the 
effective many--fermionic
operator, eq.(\ref{effop_novev}), in the bosonized theory. The correlation
function is evaluated by averaging first over the fermion fields in the
background of the pion field, later over the pion field (in saddle--point
approximation).  Let us denote the result of the averaging over the
fermions fields by $\langle \ldots \rangle_\pi$. In the leading order of
$1/N_c$ one finds
\be
\lefteqn{ \llangle \ldots \psi^\dagger (x) 
\Gamma_{\alpha_1 \ldots \alpha_r} \frac{\lambda^a}{2} 
\psi (x)
\left[ (Y_{+ \cF})^a_{\beta_1 \ldots \beta_s} (x) 
+ (Y_{- \cF})^a_{\beta_1 \ldots \beta_s}(x) \right]
\ldots  \rrangle_\pi } && \nonumber \\
&=& \frac{(i M)^{N_f}}{N_c}
\sum_{f, g = 1}^{N_f} 
\int d^4 z \, \cF_{+ , \mu\nu \beta_1 \ldots \beta_s} (x - z) 
\llangle \ldots \psi^\dagger (x) \Gamma_{\alpha_1 \ldots \alpha_r} 
\frac{\lambda^a}{2} \psi (x) 
\; J^a_{\pm\mu\nu} (z)_{fg} \ldots \rrangle_\pi \nonumber \\
&& \times \left(\frac{2V}{N}\right)^{N_f - 1} 
\llangle {\textstyle{\det'_{fg}}} J_+ (z) \rrangle_\pi \nonumber \\
&+& (+ \rightarrow -) , 
\label{corr_bosonized}
\ee
since the currents $J_\pm$ appearing in the subdeterminant are
color--singlet. (We remind that the $J_\pm$, eq.(\ref{J}), are matrices in
flavor indices.) The ellipses here stand for other functions of the fermion
fields, {\em e.g.}\ hadronic currents.  Computing the last factor in
eq.(\ref{corr_bosonized}) explicitly one finds in leading order in
$M\bar\rho$:
\be
\left( i \frac{2V}{N}\right)^{N_f - 1} 
\llangle {\textstyle{\det'_{fg}}} J_\pm (z) \rrangle_\pi 
&=& \left( i \frac{2V}{N}\right)^{N_f - 1} 
{\textstyle{\det'_{fg}}} \llangle J_\pm (z) \rrangle_\pi 
\;\; = \;\; U_\pm (z)_{fg} ,
\ee
where $U_\pm (z)$ is the pion field, eq.(\ref{U}). The first equality 
holds in leading order of $1/N_c$, and in the last step we have used the 
self--consistency condition defining the saddle point of the 
effective chiral theory. Thus, in the bosonized effective theory, 
eq.(\ref{Z_bosonized}), the instanton--induced $2 N_f$--fermionic vertex, 
eq.(\ref{Y_F}) is equivalent to the 2--fermionic ``bosonized'' vertex
\be
(Y_{\cF\pm})^a_{\beta_1 \ldots \beta_r} 
[\psi^\dagger , \psi ; \pi ]_{\rm bosonized}
&=& \frac{i M}{N_c} \int d^4 z \, 
\cF_{\pm , \mu\nu\beta_1 \ldots \beta_s} (x - z) 
\sum_{f, g}^{N_f} J^a_{\pm\mu\nu} (z)_{fg} 
U_\pm (z)_{fg} ,
\nonumber \\
\ee
which is manifestly chirally invariant. This expression formally 
includes the case $N_f = 1$, eq.(\ref{Y_F_Nf1}), if the pion field is set
to zero, $U_\pm = 1$. Accordingly, the effective operator representing 
the gluonic operator, eq.(\ref{O}), is again equivalent to a 4--fermionic 
operator,
\be
\lefteqn{\effop{O}_{\alpha_1 \ldots \alpha_r \beta_1 \ldots \beta_s} 
[\psi^\dagger , \psi ; \pi ] (x) } && \nonumber \\
&=& 
- i \psi^\dagger (x) \frac{\lambda^a}{2} \Gamma_{\alpha_1 \ldots \alpha_r} 
\psi (x) 
\left[ (Y_{\cF +})^a_{\beta_1 \ldots \beta_s} (x)_{\rm bosonized} 
+ (Y_{\cF -})^a_{\beta_1 \ldots \beta_s} (x)_{\rm bosonized} \right] ,
\label{O_eff_bosonized}
\ee
which now, however, is coupled to the pion field. This representation 
of the effective operator is useful for calculations
of nucleon matrix elements, see Section \ref{section_nucleon}. 
It is also of principal interest, since it makes manifest the chiral 
invariance of the effective operator.
\subsection{Matrix elements of higher--twist operators in quark states}
\label{subsec_quark_states}
Before computing the nucleon matrix elements of higher--twist operators it
is instructive to analyze their matrix elements in ``constituent quark''
states, {\em i.e.}, the massive quark states corresponding to the fermion
fields of the effective chiral theory, eq.(\ref{S_eff}). A pole appears in
the quark propagator after continuation to Minkowskian momenta, that is, at
Euclidean $p^2 = -M^2 + O(M\bar\rho )$.
\par
Let us first compute the quark matrix element of the effective operator
corresponding to the general gluonic operator, eq.(\ref{O}). It is
sufficient to consider one quark flavor, $N_f = 1$, in which case the
effective operator is given by the 4--fermionic operator,
eq.(\ref{effop_novev}), with the instanton--induced vertices
eq.(\ref{Y_F_Nf1}). It is convenient to carry out the entire calculation
using Euclidean fields and Euclidean vector components, and to continue to
$p^2 = -M^2$ at the end. The quark propagator in the effective theory is
given by
\be
G(k) &=& \frac{\kslash + i M F^2 (k)}{k^2 + M^2 F^4 (k)} .
\label{propagator_Nf1}
\ee
When computing the matrix elements of the 4--fermionic operator between
quark states there are two contractions (see Fig.\ref{fig_quark}).  One
obtains
\be
\lefteqn{
\langle p S | \effop{O}_{\alpha_1 \ldots \alpha_r \beta_1 \ldots \beta_s} 
| p S \rangle
\;\; = \;\; \frac{iM}{2} \fourint{k} 
\cF_{+ , \mu\nu\beta_1 \ldots \beta_s} (k ) } \nonumber \\
&\times& \left\{
\frac{F(p) F(p - k)}{(p - k)^2 + M^2 F^4 (p - k)} \;
\tr \left[ \Lambda_{p, S} \,
\Gamma_{\alpha_1 \ldots \alpha_r} 
\left( \pslash - \kslash + i M F^2(p - k) \right)
\sigma_{\mu\nu} \frac{1 + \gamma_5}{2} 
\right] \nonumber \right. \\
&&+ \left. \frac{F(p) F(p + k)}{(p + k)^2 + M^2 F^4 (p + k)} \;
\tr \left[ \Lambda_{p, S} \,
\sigma_{\mu\nu} \frac{1 + \gamma_5}{2}
\left( \pslash + \kslash + i M F^2(p + k) \right)
\Gamma_{\alpha_1 \ldots \alpha_r} \right] \right\}
\nonumber \\[.2cm]
&+& \left( \cF_{+ , \mu\nu\beta_1 \ldots \beta_s} 
\rightarrow \cF_{- , \mu\nu\beta_1 \ldots \beta_s} , \;\; 
\frac{1 + \gamma_5}{2} \rightarrow \frac{1 - \gamma_5}{2} \right) .
\label{me_quark}
\ee
It is understood that one sets $p^2 = -M^2$ after evaluating the
integrals. Here
\be
\Lambda_{p, S} &=& u (p, S) \bar u (p, S) \;\; = \;\;
\frac{-i \pslash + M}{2}
\left( 1 + i \gamma_5 \Sslash \right) 
\label{projector}
\ee
is the projector on quark states with definite momentum and
polarization vector, where $p$ and $S$ are Euclidean 4--vectors
satisfying
\be
p^2 &=& -M^2, \hspace{2em}  S^2 \; = \; 1 , 
\hspace{2em}  S\cdot p \; = \; 0.
\ee
A factor of $N_c$ resulting from the quark loop has canceled the factor
$1/N_c$ in the effective operator, eq.(\ref{Y_F_Nf1}), so that the result
is $O (N_c^0 )$. Eq.(\ref{me_quark}) is understood as the diagonal matrix
element for a quark state of given color. We remind that formula
eq.(\ref{me_quark}) is valid for operators with zero vacuum expectation
value, see the discussion in Section \ref{subsec_effective_operators}.
\par
We now apply the general formula eq.(\ref{me_quark}) to the spin--dependent
quark matrix elements of the twist--3 and 4 operators, eqs.(\ref{f2_def},
\ref{d2_def}). In the case of $f^{(2)}$ we simply contract
eq.(\ref{f2_def}) with the polarization vector, $S$; to calculate $d^{(2)}$
we contract eq.(\ref{d2_def}) with a light--like vector, $n$, which we
choose orthogonal to the polarization vector in order to put the trace
terms in eq.(\ref{d2_def}) to zero,
\be
n^2 \; = \; 0, \hspace{2cm} n\cdot S \; = \; 0 .
\ee
Passing to Euclidean fields and vector components we have
\be
2 M^3 f^{(2)}_{\rm quark}
&=& - S_\beta \; \langle p S | \bar\psi \gamma_\alpha 
\Fdual_{\beta\alpha} \psi 
| p S \rangle , \label{f2_projected} \\
4 M^3 d^{(2)}_{\rm quark} &=& n_\alpha n_\beta S_\gamma \;
\langle p S | \bar\psi \gamma_\alpha \Fdual_{\beta\gamma} \psi 
| p S \rangle .
\label{d2_projected}
\ee
To evaluate this matrix element we need the Fourier transform of the 
dual of the field strength of one $I$ and $\bar I$, eq.(\ref{Fdual_inst}), 
which is given by
\be
\lefteqn{
\pm \int d^4 x \, \exp (-ik\cdot x) 
\frac{8  \rho^2 }{(x^2 + \rho^2 )^2}
\left( \frac{x_\mu x_\beta}{x^2} \delta_{\gamma\nu} 
+ \frac{x_\nu x_\gamma}{x^2} \delta_{\mu\beta} 
- \half \delta_{\mu\beta} \delta_{\gamma\nu} \right) } && \nonumber \\
&=& \pm \rho^2 {\cal G} (k) 
\left( \frac{k_\mu k_\beta}{k^2} \delta_{\gamma\nu} 
+ \frac{k_\nu k_\gamma}{k^2} \delta_{\mu\beta} 
- \half \delta_{\mu\beta} \delta_{\gamma\nu} \right) ,
\label{F_inst_four}
\ee
\be
{\cal G} (k) &=& 32 \pi^2 \left[ \left(\frac{1}{2} + \frac{4}{t^2} \right) 
K_0 (t) + \left( \frac{2}{t} + \frac{8}{t^3} \right) K_1 (t) 
- \frac{8}{t^4} \right], \hspace{1cm} t \; = \; k\rho , 
\label{G} 
\ee
where $K_n (t)$ are modified Bessel functions of the second kind. Inserting
this in eq.(\ref{me_quark}), adding the contributions from $I$'s and 
$\bar I$'s, we obtain
\be
2 M^3 f^{(2)}_{\rm quark} &=& 3 M \bar\rho^{-2} \; I_1 (p) ,
\nonumber \\
I_1 (p) &=& \bar\rho^4 \fourint{k} 
\frac{{\cal G} (k) F(p) F(p - k)}{(p - k)^2 + M^2 F^4 (p - k)} \, k\cdot p ,
\label{f2_integral}
\ee
and
\be
4 M^3 d^{(2)}_{\rm quark} &=& -2 M^3 \; I_2 (p) ,
\nonumber \\
I_2 (p) &=& \bar\rho^{-2} \fourint{k} 
\frac{{\cal G}(k) F(p) F^3 (p - k)}{(p - k)^2 + M^2 F^4 (p - k)} 
\frac{1}{3} \left( - 1 + 4 \frac{(k\cdot p)^2}{k^2 p^2} \right) ,
\label{d2_integral}
\ee
where one should set $p^2 = -M^2$ after evaluating the integrals.  One
observes that $f^{(2)}_{\rm quark}$ and $d^{(2)}_{\rm quark}$ are of 
different order in the
parameter $M\bar\rho$. For small $p$ both integrals $I_1 (p)$ and $I_2 (p)$
are proportional to $\bar\rho^2 p^2$, and thus
\be
f^{(2)}_{\rm quark} &\sim& (M\bar\rho )^0 , \nonumber \\
d^{(2)}_{\rm quark} &\sim& (M\bar\rho )^2 \log M\bar\rho 
\label{f2_d2_parametric}
\ee
for $M\bar\rho \rightarrow 0$. Numerically one finds, evaluating the
integrals for small $p$,
\be
I_1 (p) &=& - \frac{2}{5} \; \bar\rho^2 p^2 , 
\nonumber \\
f^{(2)}_{\rm quark} &=& \frac{3}{5} \;\; = \;\; 0.6
\label{f2_quark}
\ee
in the limit $M\bar\rho \rightarrow 0$, and
\be
I_2 (p) &=& -0.066 \; \bar\rho^2 p^2 , \nonumber \\
d^{(2)}_{\rm quark} &=& -0.011
\label{d2_quark}
\ee 
at the phenomenological value $M\bar\rho = 0.58$. The different parametric
order of the twist--4 and twist--3 matrix elements reflects itself in the
numerical values: $d^{(2)}_{\rm quark}$ is more than an order of magnitude
smaller than $f^{(2)}_{\rm quark}$.
\par
Similarly, we can apply the general formula eq.(\ref{me_quark}) to the case
of the spin--in\-de\-pen\-dent 
twist--4 operator contributing to power corrections 
to the Gross--Llewellyn-Smith sum rule, eq.(\ref{c2_def}).  Contracting
eq.(\ref{c2_def}) with the 4--momentum vector and passing to Euclidean 
fields and vector components we have
\be
2 M^4 c^{(2)}_{\rm quark} &=& p_\beta \; \half \sum_{\rm spin}
\langle p | \bar\psi \gamma_\alpha \gamma_5 
\Fdual_{\beta\alpha} \psi | p \rangle .
\label{c2_projected}
\ee
Evaluating the matrix element with the help of eq.(\ref{me_quark}) and
summing over spins we obtain
\be
2 M^4 c^{(2)}_{\rm quark} &=& -3 M^2 \; I_1 (p).
\label{c2_integral}
\ee
Thus, 
\be
c^{(2)}_{\rm quark} &\sim&  (M\bar\rho )^0 ,
\label{c2_parametric}
\ee
just as the spin--dependent twist--4 matrix element, $f^{(2)}_{\rm quark}$,
eq.(\ref{f2_d2_parametric}), and numerically in the limit 
$M\bar\rho \rightarrow 0$
\be
c^{(2)}_{\rm quark} &=& -\frac{3}{5} \;\; = \;\; -0.6 .
\label{c2_quark}
\ee
\par
Finally, we note that an analogous calculation of the forward matrix
elements of the operators
\be
\langle p | \bar\psi \gamma_\alpha F_{\beta\alpha}
\psi | p \rangle , \hspace{2em}
\langle p S | \bar\psi \gamma_\alpha \gamma_5 F_{\beta\alpha}
\psi | p S \rangle ,
\ee
gives identically zero, as it should be on grounds of the QCD equations of
motion. From this we conclude that the method of effective operators
preserves a principal feature of QCD: matrix elements of operators which
are zero in QCD due to the QCD equations of motion are automatically zero
in the effective theory.  This remarkable property of effective operators
will be investigated in more detail in Section \ref{subsec_eom}.
\par
The results obtained here for the quark matrix elements of higher--twist
operators can be rephrased in terms of local operators, whose matrix
elements reproduce those of the non-local effective 4--fermionic operators
at quark level. At momenta of order $M \ll \bar\rho^{-1}$ this
``localization'' is justified by the fact that the effects of the
non-locality of the instanton--induced vertices contribute only in higher
orders of $M\bar\rho$. Since the quark loop integrals in
eqs.(\ref{f2_integral}, \ref{d2_integral}, \ref{c2_integral}) are
proportional to the external momentum $p^2$, the local operator which
should be valid for general quark momenta $p$ of order $M$, not necessarily
on the mass shell $p^2 = -M^2$, must contain derivatives.  The twist--4
operator, eq.(\ref{f2_def}), we can thus represent as (in the Euclidean
theory)
\be
\bar\psi \gamma_\alpha \Fdual_{\beta\alpha} \psi  
&\rightarrow &
f^{(2)}_{\rm quark} \; \bar\psi \gamma_\beta \gamma_5 \partial^2 \psi .
\label{f2_equiv}
\ee
Up to the derivative the equivalent local operator is the axial current
operator, multiplied by the coefficient $f^{(2)}_{\rm quark}$.  For the
twist--3 operator, eq.(\ref{d2_def}), the corresponding local operator is
\be
\bar\psi \left(\gamma_\alpha \Fdual_{\beta\gamma} 
+ \gamma_\beta \Fdual_{\alpha\gamma}  \right) \psi  
& \rightarrow  &
- d^{(2)}_{\rm quark} \; \bar\psi
\left( \partial_\alpha \partial_{\left[ \beta \right.} 
\gamma_{\left. \gamma \right]} 
+ \partial_\beta \partial_{\left[ \alpha \right.} 
\gamma_{\left. \gamma \right]} \right)
\gamma_5 \psi  .
\label{d2_equiv}
\ee
For the spin--independent twist--4 operator, eq.(\ref{c2_def}), 
finally, the equivalent local operator is related to the vector current,
\be
\bar\psi \gamma_\alpha \gamma_5 \Fdual_{\beta\alpha} 
\psi  
&\rightarrow & 
c^{(2)}_{\rm quark} \; \bar\psi \gamma_\beta \partial^2 \psi  .
\label{c2_equiv}
\ee
When going to nucleon, for various reasons, a strictly local representation
of the higher--twist operators is no longer possible.  In the nucleon one
finds ``sea'' quarks with momenta of order of the cutoff, $\bar\rho^{-1}$,
so one must take into account the full momentum dependence of the
instanton--induced vertices. Furthermore, there are interaction
contributions to the matrix element in which more than one quark connects
to the many--fermionic operator. The local operators are nevertheless
useful as a replacement for the non-local effective operator for
``valence'' quarks whose momenta are restricted to values 
$\ll \bar\rho^{-1}$ by the bound--state wave function of the nucleon.
Also, the parametric order in $M\bar\rho$ which is encoded in the constants
$f^{(2)}_{\rm quark}, d^{(2)}_{\rm quark}$ and $c^{(2)}_{\rm quark}$ will
carry over to the nucleon matrix elements.  In Section
\ref{subsec_nucleon_leading} we shall see to which extent the relation
between $f^{(2)}$ and the axial vector coupling constant, $g_A$, as well as
between $c^{(2)}$ and the vector coupling constant, which holds for
on-shell quark matrix elements, is valid also for the nucleon.
\subsection{The QCD equation of motion}
\label{subsec_eom}
The set of local operators which are used in the operator product expansion
at higher--twist level is overcomplete. Some higher--twist operators can 
be reduced to other operators, others are identically zero, by the QCD 
equations of motion. For instance, the twist--4 operator
\be
\bar\psi (x) \nablaslash \psi (x)
&=& \bar\psi (x) \gamma_\alpha 
\left( \partial_\alpha - i \frac{\lambda^a}{2} A_\alpha^a (x) \right) 
\psi (x)
\label{eom_operator}
\ee 
should have zero matrix element in physical states,
\be
\langle P | \psi^\dagger \nablaslash \psi | P \rangle 
&=& 0. 
\label{eom}
\ee
Furthermore, using the fact that in QCD
\be
\frac{\lambda^a}{2} F_{\beta\alpha}^a (x) &=& i [\nabla_\beta , 
\nabla_\alpha ],
\ee
one may show that
\be
\langle P | \bar\psi \gamma_\alpha F_{\beta\alpha} \psi | P \rangle 
&=& 0 .
\label{F_gamma_null}
\ee
When passing from QCD to the effective theory it is important that 
these general properties are preserved by the corresponding effective
operators. We already noted in Section \ref{subsec_quark_states}
that the matrix element eq.(\ref{F_gamma_null}) in quark states
is identically zero. We now want to show in a more general context
that eq.(\ref{eom}) holds in the effective theory, when the gauge field 
in the QCD operator, eq.(\ref{eom_operator}), is replaced by the 
corresponding effective operator.
\par
Consider the gauge field part of the QCD operator eq.(\ref{eom_operator}).
Since we are interested in the forward matrix element we may integrate 
the operator over the 4--volume,
\be
O_A &=& \int d^4 x \left[ \bar\psi (x) \gamma_\alpha
\frac{\lambda^a}{2} \psi (x) A_\alpha^a (x) \right] .
\ee
The effective operator corresponding to this operator is given 
by eq.(\ref{effop})\footnote{Since the instanton--induced vertices 
$(Y_{A \pm})^a_\alpha (x)$ can have a non-zero vacuum expectation value 
we must use here the full form of the effective operator, including 
the factors $\cR_\pm$.}
\be
\effop{O_A} &=& \int d^4 x \left[
- i \psi^\dagger (x) \gamma_\alpha \frac{\lambda^a}{2} \psi (x) 
\left\{ (Y_{A +})^a_\alpha (x) N_+ \cR_+ + 
(Y_{A -})^a_\alpha (x) N_- \cR_-  \right\} \right] .
\label{Aslash_eff}
\ee
Here the vertices $(Y_{A \pm})^a_\alpha$ are given by the general formula
eq.(\ref{Y_F}), in which one substitutes the gauge potential 
of one $I$ ($\bar I$), eq.(\ref{A_inst}), the Fourier transform
of which is given by
\be
\int d^4 x \, e^{-i k\cdot x}
\frac{2 \rho^2}{(x^2 + \rho^2 ) x^2} x_\nu \delta_{\mu\beta} 
&=& -i \rho^4 {\cal H}(k) k_\nu \delta_{\mu\beta} ,
\ee
\be
{\cal H}(k) &=& \frac{8 \pi^2}{t^2} \left[ - \frac{1}{2} K_0 (t)
- \frac{2}{t} K_1 (t) + \frac{2}{t^2} \right] ,
\hspace{1cm} t \; = \; k\rho .
\label{A_inst_four}
\ee
Consider a correlation function (connected average) of the effective 
operator eq.(\ref{Aslash_eff}) with external fermion fields. The 
connected average consists of two parts, in which the external 
fermion fields connect either to the $2 N_f$--fermionic vertices, 
$(Y_{A \pm})^a_\alpha$, or to the factors $\cR_\pm$: 
\be
\lefteqn{
\llangle \psi (y' ) \; \effop{O_A} \; \psi^\dagger (y)
\rrangle_{\rm conn} } && \nonumber \\
&=& 
\llangle \psi (y' ) \int d^4 x \left[ -i \psi^\dagger (x ) \gamma_\mu 
\frac{\lambda^a}{2} \psi (x ) (Y_{A +})^a_\mu (x) \right] 
\psi^\dagger (y)
\rrangle_{\rm conn} N_+ \llangle \cR_+ \rrangle \nonumber \\
&+&
\llangle \int d^4 x \left[ - i \psi^\dagger (x ) \gamma_\mu 
\frac{\lambda^a}{2} \psi (x ) (Y_{A +})^a_\mu (x ) \right]
\rrangle N_+ \llangle \psi (y' ) \cR_+ \psi^\dagger (y) 
\rrangle_{\rm conn} \nonumber \\
&+& (+ \rightarrow -).
\label{Aslash_corr}
\ee
By explicit calculation, using eqs.(\ref{Y_F_Nf1}, \ref{A_inst_four}) 
and the fact that the form factor $F(k)$ is proportional to the Fourier 
transform of the instanton zero mode, which implies the relation
\be
\fourint{k} \; {\cal H} (p - k) \; \frac{(p - k)\cdot k}{k^2} \; F(k)
&=& \frac{2i}{3} F(p) ,
\ee
one shows that up to terms of order $(M\bar\rho)^2$ the insertion of the
$(2 N_f + 2)$--fermionic vertex into the correlation function of fermion
fields is equivalent to an insertion of a 't Hooft vertex, eq.(\ref{Y}):
\be
\llangle \psi (y' ) \int d^4 x \left[ -i \psi^\dagger (x) \gamma_\mu 
\frac{\lambda^a}{2} \psi (x) \; (Y_{A \pm})^a_\mu (x) \right] 
\psi^\dagger (y)
\rrangle_{\rm conn} 
&=&
2 \llangle \psi (y' ) \; Y_\pm \; \psi^\dagger (y) .
\rrangle_{\rm conn}
\nonumber \\
\ee
The factor of $2$ appears because, due to the color structure, there are 
$2 N_f$ different ways to contract the $(2 N_f + 2)$--fermionic 
operator on the right--hand side with the external quark fields, 
but only $N_f$ for the $2 N_f$--fermionic vertices $Y_\pm$. 
On the other hand, 
the vacuum expectation value of the $(2 N_f + 2)$--fermionic vertex 
appearing in the second term of eq.(\ref{Aslash_corr}) is
\be
\llangle \int d^4 x \left[
- i \psi^\dagger (x ) \gamma_\mu 
\frac{\lambda^a}{2} \psi (x ) \; (Y_{A \pm})^a_\mu (x )\right]
\rrangle
&=&
\llangle Y_\pm \rrangle \;\; = \;\; N_\pm
\ee
by virtue of the self--consistency condition defining the saddle point.
The connected average of the factors $\cR_\pm$ with external fermion 
fields is given by eq.(\ref{R_conn}), which is a consequence of their
definition as the ``inverse'' of the 't Hooft vertices.
Putting these results together, one observes that the sum of the 
two terms of eq.(\ref{Aslash_corr}) combines to give
\be
\llangle \psi (y' ) \; \effop{O_A} \; \psi^\dagger (y) \rrangle_{\rm conn}  
&=& \llangle \psi (y' ) \; Y_\pm \; \psi^\dagger (y) \rrangle_{\rm conn} .
\ee
We have thus shown that, to order $(M\bar\rho )^2$, the gluonic part 
of the QCD equations of motion, eq.(\ref{eom}), corresponds to an insertion 
of the 't Hooft vertex in the effective theory. The operator corresponding 
to the QCD equations of motion thus vanishes by virtue of the equations
of motion of the effective theory, a very gratifying result. 
This serves as additional confirmation that the approximations made in 
deriving the effective action and the effective operators are mutually
consistent.
\subsection{Non-leading vs. leading twist}
\label{subsection_twist}
In section \ref{subsec_quark_states} it was shown that matrix elements of
gluonic operators of twist 4 are of order unity in the packing fraction
of the instanton medium, while those of twist 3 are suppressed.
We now want to investigate the role of instantons in operators
of twist 2, which determine the scaling (non--power suppressed) part
of nucleon structure functions. In particular, using the method of
effective operators, we want to see why --- and under which conditions ---
it is justified to drop the gluon field in twist--2 QCD operators
when passing to the effective chiral theory. This approximation of
``quarks--antiquarks only'' is used in all calculations of 
leading--twist structure functions of the nucleon in the effective 
chiral theory \cite{DPPPW96,DPPPW97}.
\par
Consider the twist--2 QCD operator of spin $n + 1$, eq.(\ref{a2_S}), which 
determines the $n$'th moment of the polarized structure function, $g_1$
(for simplicity let us consider one flavor, $N_f = 1$). With the 
explicit form of the covariant derivative,
\be
\nabla_\mu &=& \partial_\mu - i \frac{\lambda^a}{2} A_\mu^a (x) ,
\ee
the operator can be written as a part containing only explicit derivatives,
plus terms involving powers of the gauge field\footnote{We remind that we 
have in mind here the gauge in which the instanton field 
takes the form eq.(\ref{A_inst}), the so-called singular gauge.},
\be
\bar\psi \gamma_{\left\{\alpha \right.} \gamma_5 
\nabla_{\beta_1} \ldots \nabla_{\left. \beta_n \right\}} \psi  
&=& 
\bar\psi \gamma_{\left\{\alpha \right.} \gamma_5 
\partial_{\beta_1} \ldots \partial_{\left. \beta_n \right\}} \psi  
\nonumber \\
&-& i\sum_i^n
\bar\psi \gamma_{\left\{\alpha \right.} \gamma_5 
\partial_{\beta_1} \ldots A^a_{\beta_i}
\ldots \partial_{\left. \beta_n \right\}} \frac{\lambda^a}{2} \psi  
\nonumber \\
&+& \mbox{terms} \;\; A^2, A^3 \ldots
\label{twist2_explicit}
\ee
When passing to the effective chiral theory, the pure derivative part
becomes simply the corresponding operator in terms
of fields of the effective theory, while the terms involving the gauge
field have to be replaced by the corresponding effective operators;
they are of the form of the general operator eq.(\ref{O}).
Let us consider the matrix elements of eq.(\ref{twist2_explicit})
in quark states of the effective theory. The matrix element of the pure 
derivative part is trivial and contributes precisely unity to 
the moment $a^{(n)}$. On the other hand, using the general formulas
of sections \ref{subsec_effective_operators} and 
\ref{subsec_quark_states} one easily shows that the matrix element of 
the terms involving the gauge field is of order $(M\bar\rho )^2$.
Thus,
\be
a^{(n)}_{\rm quark} &=& 1 \; + \; O\left( (M\bar\rho )^2 \right) .
\label{an_quark}
\ee
The contribution of the gauge field is parametrically suppressed
relative to the pure derivatives. In terms of the structure function
$g_1$ this implies
\be
g_1 (x)_{\rm quark} &=& \half \delta (x - 1) \; + \; 
O\left( (M\bar\rho )^2 \right) .
\ee
To leading order in $M\bar\rho$ the quark of the effective chiral theory
is ``elementary'', it structure in the instanton vacuum emerges only at 
level $(M\bar\rho )^2$. 
\par
At the same time, one finds that the twist--2 QCD operators determining
the moments of the polarized gluon distribution,
\be
\lefteqn{
\langle P S | 
F_{\left\{\alpha \beta \right.} D_{\beta_1} \ldots 
D_{\beta_{n - 1}} \Fdual^\beta_{\left.\beta_n \right\}} 
| P S \rangle - \mbox{traces} } && \nonumber \\
&=& 2 M_N \; b^{(n)} \;
S_{\left\{\alpha \right.} P_{\beta_1} \ldots P_{\left. \beta_n \right\}} 
- \mbox{traces} ,
\label{twist2_gluon}
\ee
are identically zero when one substitutes the field of one instanton
(the same applies to the unpolarized gluon distribution.)
The reason is the $O(4)$ symmetry of the instanton field: after integration 
over instanton coordinates one can build the $\beta_1\ldots\beta_n$ tensor 
only out of Kronecker symbols, so it is impossible to get it traceless. 
Thus the effective operator for eq.(\ref{twist2_gluon}) is zero.
However, the operator in eq.(\ref{twist2_gluon}) becomes non-zero when 
taking into account two--instanton contributions, which are
parametrically suppressed by a factor 
$(\bar\rho / R )^4 \propto (M\bar\rho)^2$.
We conclude that
\be
b^{(n)} &=& O\left( (M\bar\rho )^2 \right) .
\label{gn_quark}
\ee
\par
To summarize, we see that {\em in twist--2 operators the effect of the 
instanton field is always of order} $(M\bar\rho )^2$. When working to 
leading order in $M\bar\rho$, it is consistent to keep only the pure 
derivative part in the twist--2 quark operators, eq.(\ref{twist2_explicit}), 
and to replace the operators for the twist--2 gluon distribution
by zero. This is the approximation employed in the calculation of the
twist--2 part of the nucleon structure functions in 
refs.\cite{DPPPW96,DPPPW97,PP96}.
\par
However, as we have seen above, {\em instantons can make contributions 
of order unity in higher--twist operators}. In the cases considered in 
section \ref{subsec_quark_states} the first non-suppressed instanton 
contributions appeared at twist 4. There can of course also be 
$O((M\bar\rho )^2 )$ contributions to operators of twist 6 and higher.
\par
The difference between the instanton contribution to operators of leading 
and non-leading twist can be seen even more clearly in the case 
of unpolarized structure functions. The second moment of the
structure function $F_2$, $M^{(2)}$, is given by the matrix element of 
the twist--2 spin--2 operator,
\be
\langle P | \bar\psi \gamma_{\left\{\mu_1 \right.}
\nabla_{\left. \mu_2 \right\} } \psi | P \rangle - \mbox{traces} 
&=& M^{(2)} P_{\left\{\mu_1 \right.} P_{\left. \mu_2 \right\}} 
- \mbox{traces} .
\label{M2}
\ee
In analogy to eq.(\ref{an_quark}), at quark level the pure derivative 
term makes unity contribution to $M^{(2)}$, while the contribution
of the gauge field is suppressed by $(M\bar\rho )^2$:
\be
M^{(2)}_{\rm quark} &=& 1 \; + \; O\left( (M\bar\rho )^2 \right) 
\ee
This is consistent with the energy--momentum sum rule, since 
the momentum carried by gluons is $O( (M\bar\rho )^2 )$, 
{\em cf.}\ eq.(\ref{gn_quark}). On the other hand, if one does not 
project on twist 2 but contracts the indices in eq.(\ref{M2})
one obtains the twist--4 matrix element
\be
\langle P | \bar\psi \nablaslash \psi | P \rangle &=& C M^2 ,
\ee
where $C = 0$ by virtue of the equations of motion. As was is shown in 
section \ref{subsec_eom}, this zero comes because an order 
unity contribution to $C$ from the instanton field exactly cancels the
pure derivative term in the Dirac operator.
\section{Nucleon matrix elements of higher--twist operators}
\setcounter{equation}{0}
\label{section_nucleon}
\subsection{The nucleon in the effective chiral theory}
An immediate application of the effective chiral theory derived from
the instanton vacuum is the chiral quark soliton model of the 
nucleon \cite{DPP88}. In this section we briefly recall the 
main characteristics of this description.
\par
In the effective chiral theory the nucleon is described in the large--$N_c$ 
limit by a classical pion field, $U_c ({\bf x} )$. 
The quarks single-particle wave functions are determined from 
the Dirac equation in the external pion field,
\be
H \Phi_n ({\bf x}) &=& E_n\Phi_n ({\bf x}),
\label{dirac}
\ee
\be
H &=& \gamma_4 \left[ \gamma_k \partial_k - M 
U^{\gamma_5} ({\bf x}) \right] ,
\label{H} \\
U^{\gamma_5} &=& \exp\left[ i\pi^a(x)\tau^a\gamma_5\right] 
\;\; = \;\; \frac{1 + \gamma_5}2 U_+ (x) + 
\frac{1 - \gamma_5}2 U^\dagger (x) .
\ee
The spectrum of the one-particle Hamiltonian, $H$, contains the upper and 
lower Dirac continua (distorted by the presence of the external pion field), 
as well as a discrete bound--state level. We denote the energy of this 
discrete level by $E_{\rm lev},\;\;-M \leq E_{\rm lev} \leq M$. This level 
must be occupied by $N_c$ quarks to have a state of unit baryon number. 
The saddle point pion field is determined by minimizing the static energy, 
which includes the energy of the discrete level as well as the aggregate 
energy of the negative Dirac continuum. It is of ``hedgehog'' form,
\be
U_c ({\bf x}) \; = \; \exp\left[i\pi^a({\bf x})\tau^a\right]
\; = \; \exp\left[\frac{ i {\bf x} \cdot \bftau }{r} P(r)\right],
\hspace{1cm} r \; = \; |{\bf x}|,
\label{hedge}
\ee
where $P(r)$ is called the profile function. A variational approximation
to the soliton profile in the form 
\be
P(r) &=& 2\;\arctan\left(\frac{r_0^2}{r^2}\right)
\label{varprof}
\ee
with $r_0 \approx 1.0/M$ gives a very reasonable description of a variety
of static nucleon observables \cite{DPP88}.
\par
The minimum of the energy is degenerate with respect to
translations of the soliton in space and to rotations of the soliton
field in ordinary and isospin space. For the hedgehog field eq.(\ref{hedge}) 
the two rotations are equivalent. Quantizing slow rotations of the saddle-point
pion field gives rise to the quantum numbers of the nucleon: its spin and
isospin components \cite{ANW,DPP88}. In order to take into account the
translational and rotational zero modes one has to perform a rotation 
of the soliton field in flavor space and to shift its center,
\be
U({\bf x}) &=& R U_c({\bf x} - {\bf X}) R^\dagger , 
\label{rotation_U}
\ee
the same for the quark eigenfunctions,
\be
\Phi_n ({\bf x}) &\rightarrow& R \Phi_n({\bf x} - {\bf X}),
\label{rotation_Phi}
\ee
and to make a projection to a concrete nucleon state under
consideration. The projection on a nucleon state with given momentum
${\bf P}$ is obtained by integrating over all shifts ${\bf X}$ of the
soliton,
\be
\langle {\bf P^\prime}|\ldots|\ {\bf P}\rangle
&=& \int d^3{\bf X}\;e^{i({\bf P^\prime-P})\cdot{\bf X}}\; \ldots .
\label{totmom}
\ee
The projection on a nucleon with given spin ($S_3$) and isospin
($T_3$) components is obtained by integrating over all spin-isospin
rotations $R$ of the soliton,
\be
\langle S=T,S_3,T_3|\ldots| S=T,S_3,T_3\rangle
&=& \int dR\;\phi^{\dagger\;S=T}_{S_3T_3}(R) \; \ldots \; 
\phi^{S=T}_{S_3T_3}(R),
\label{spisosp}
\ee
where $\phi^{S=T}_{S_3T_3}(R)$ is the rotational wave function of the
nucleon given by the Wigner finite-rotation matrix \cite{DPP88}:
\be
\phi^{S=T}_{S_3T_3}(R) &=&
\sqrt{2S+1}(-1)^{T+T_3}D^{S=T}_{-T_3,S_3}(R).
\label{Wigner}
\ee
The four nucleon states have $S=T=1/2$, with $S_3,T_3=\pm 1/2$.
It is implied that $dR$ in eq.(\ref{spisosp}) is the Haar measure normalized 
to unity. (In the following we shall omit the superscript $S=T$.)
\subsection{Nucleon matrix elements of effective gluon operators}
\label{subsec_nucleon_effective}
We now turn to the calculation of nucleon matrix elements of the
effective operators which arise in the description of higher--twist
corrections. In this subsection we derive the expressions for the
nucleon matrix element of the general gluonic operator, eq.(\ref{O}), 
which was considered in Sections \ref{subsec_effective_operators}
and \ref{subsec_quark_states},
\be
O_{\alpha_1 \ldots \alpha_r \beta_1 \ldots \beta_s} (x) 
&=& \bar\psi (x) \; \hat{t} \; \frac{\lambda^a}{2} 
\Gamma_{\alpha_1 \ldots \alpha_r} \, \psi (x) \;
\cF^a_{\beta_1 \ldots \beta_s} (x) ,
\nonumber
\ee
where we have introduced now a flavor matrix, $\hat{t}$, with
$\hat{t} = \tau^3$ for the flavor--nonsinglet and $\hat{t} = 1$ 
for the singlet case.
\par
To compute the nucleon matrix element we use the bosonized form of the 
effective operator, eq.(\ref{O_eff_bosonized}), where the pion field 
is to be replaced by the soliton field, eq.(\ref{hedge}):
\be
\lefteqn{\effop{O}_{\alpha_1 \ldots \alpha_r \beta_1 \ldots \beta_s} (x)}
&& \nonumber \\
&=& \frac{i M}{N_c} \int d^4 z \, 
\cF_{+ , \mu\nu \beta_1 \ldots \beta_s} (x - z) 
\nonumber \\
&&\times 
\left[ - i \psi^\dagger (x) \; \hat{t} \; \frac{\lambda^a}{2} 
\Gamma_{\alpha_1 \ldots \alpha_r} 
\psi (x) \right] \; \left[
\psi^\dagger (z) \frac{\lambda^a}{2} \sigma_{\mu\nu} U_+ ({\bf z})
\frac{1 + \gamma_5}{2} \psi (z) \right]
\nonumber \\[.2cm]
&+& \left( \cF_{+ , \mu\nu\beta_1 \ldots \beta_s} 
\rightarrow \cF_{- , \mu\nu\beta_1 \ldots \beta_s} , \;\; 
\frac{1 + \gamma_5}{2} \rightarrow \frac{1 - \gamma_5}{2} , \;\; 
U_+ \rightarrow U_- \right) .
\label{effop_explicit}
\ee
(For simplicity we have neglected for the moment the nonlocality of the 
instanton--induced vertices, {\em i.e.}, we have set the form 
factor $F(k) \equiv 1$. The form factors will be inserted again later.)
This operator has the form of a product of two
color--octet currents, similar to one--gluon exchange between 
quarks, with the functions $\cF_{\pm , \mu\nu\beta_1 \ldots \beta_s}$ 
playing the role of the gluon propagator. The calculation of the nucleon 
matrix element of the effective
operator is thus analogous to that of gluon exchange corrections to the 
nucleon mass \cite{Jaenicke91}. One expands the quark fields 
in eq.(\ref{O_eff_bosonized}) in the basis of single--particle wave 
functions, eq.(\ref{dirac}), and computes the matrix element, summing 
over all occupied quark single--particle levels including the bound--state 
level and the negative Dirac continuum.
In addition, one has to perform a rotation and shift of the center 
of the pion field and the quark wave functions, 
eqs.(\ref{rotation_U} , \ref{rotation_Phi}), and integrate over 
collective wave functions to project on a nucleon state
with given spin--isospin quantum numbers. One obtains
\be
\lefteqn{ \langle {\bf P} = 0, T_3, S_3 | 
\effop{O}_{\alpha_1 \ldots \alpha_r \beta_1 \ldots \beta_s} 
| {\bf P} = 0, T_3, S_3 \rangle } && \nonumber \\
&=& 2 M_N \int dR \, \phi_{S_3 T_3}^\dagger(R) \phi_{S_3 T_3} (R) 
\; {\cal M}_{\alpha_1 \ldots \alpha_r \beta_1 \ldots \beta_s}(R) ,
\ee
\be
\lefteqn{ {\cal M}_{\alpha_1 \ldots \alpha_r \beta_1 \ldots \beta_s}(R) } 
&& \nonumber \\
&=& - \frac{N_c}{2 T} 
\int d^4 x \int d^4 z \; \cF_{+ , \mu\nu\beta_1 \ldots \beta_s} (x - z) \;
{\sum_{m, n}}'
\exp \left[ (E_n - E_m ) (x_4 - z_4 ) \right] 
\nonumber \\
&&\times 
\left[ \Phi_n^\dagger ({\bf x}) \gamma_4 \; R^\dagger \hat{t} R \;
\Gamma_{\alpha_1 \ldots \alpha_r} 
\Phi_m ({\bf x}) \right]
\left[ \Phi_m^\dagger ({\bf z}) \gamma_4 \; \sigma_{\mu\nu} 
\frac{1 + \gamma_5}{2} M U_+ ({\bf z}) \;  
\Phi_n ({\bf z}) \right]
\nonumber \\
&+& (+ \rightarrow - ) ,
\label{O_general_levels}
\ee
where
\be
{\sum_{m, n}}' &=& \theta (x_4 - z_4 ) 
\sum\limits_{\scriptstyle n\atop \scriptstyle{\rm occ.}}
\sum\limits_{\scriptstyle m\atop \scriptstyle{\rm non-occ.}}
\; + \; \theta (z_4 - x_4 ) 
\sum\limits_{\scriptstyle n\atop \scriptstyle{\rm non-occ.}}
\sum\limits_{\scriptstyle m\atop \scriptstyle{\rm occ.}} .
\ee
Here we have used that for the forward matrix 
element, ${\bf P} = {\bf P}' = 0$, the integral over translations
is trivial. Due to translational invariance it is equivalent to 
integrating the coordinate of the operator, $x$, eq.(\ref{O_eff_bosonized}), 
over the 3--volume. To obtain a more symmetric form, we have integrated 
over the Euclidean 4--volume and divided by the Euclidean time 
interval, $T$. In the following we shall drop the labels ${\bf P} = 0$.
\par
A factor of $N_c^2 / 2$ resulting from the sum over color indices 
combines with the factor $1/N_c$ in the effective operator 
eq.(\ref{O_eff_bosonized}), to give an overall factor of $N_c/2$.
\par
To proceed with the evaluation of eq.(\ref{O_general_levels}) one has 
to perform the integral over soliton rotations,
considering separately flavor singlet and nonsinglet operators.  
We note that the factor $\Phi^\dagger_m \ldots U_\pm \ldots \Phi_n$
in eq.(\ref{O_general_levels}) does not depend on the rotation matrices, 
as it should be. This factor represents the ``gluonic'' part of the 
operator eq.(\ref{effop_explicit}), and is thus by definition a flavor 
singlet.
\par
One easily sees that for spin--dependent matrix elements such as 
$d^{(2)}$ and $f^{(2)}$, eqs.(\ref{d2_def}, \ref{f2_def}), 
the flavor--nonsinglet part appears in the leading order of 
the $1/N_c$--expansion. In this case in eq.(\ref{O_general_levels}) 
$\hat{t} = \tau^3$, and the integral over rotations is performed using
\be
R^\dagger \tau^3 R &=& D_{3a} (R) \tau^a ,
\hspace{2em}
D_{ab}(R) \;\; = \;\; \half \,
\tr \left[ R^\dagger\tau^a R \tau^b \right],
\ee
\be
\int dR \, \phi_{S_3 T_3}^\dagger (R) D_{3a} (R)
\phi_{S_3 T_3} (R)  &=& -\frac{1}{3} (\tau^3 )_{T_3 T_3} 
(\sigma^a )_{S_3 S_3} .
\label{int_rot_isovector}
\ee
The non-singlet matrix element thus becomes
\be
\langle T_3, S_3 | 
\effop{O}^{NS}_{\alpha_1 \ldots \alpha_r \beta_1 \ldots \beta_s} 
| T_3, S_3 \rangle_{\rm spin-dep.}
&=& - 2 M_N (2 T_3) \frac{S_a}{3} \; 
{\cal M}^a_{\alpha_1 \ldots \alpha_r \beta_1 \ldots \beta_s} ,
\nonumber 
\ee
\be
{\cal M}^a_{\alpha_1 \ldots \alpha_r \beta_1 \ldots \beta_s}
&=& -\frac{N_c}{2 T}
\int d^4 x \int d^4 z \cF_{+ , \mu\nu \beta_1 \ldots \beta_s} (x - z) 
{\sum_{m, n}}' \exp \left[ (E_n - E_m ) (x_4 - z_4 ) \right] 
\nonumber \\
&&\times \left[ \Phi_n^\dagger ({\bf x}) \gamma_4 \tau^a 
\Gamma_{\alpha_1 \ldots \alpha_r} \Phi_m ({\bf x}) \right]
\left[ \Phi_m^\dagger ({\bf z}) \gamma_4 \sigma_{\mu\nu} 
\frac{1 + \gamma_5}{2} M U_+ ({\bf z})  
\Phi_n ({\bf z}) \right]
\nonumber \\
&+& (+ \rightarrow - ) .
\label{O_NS_levels}
\ee
Here $S_a\; (a = 1, 2, 3)$ are the components of the nucleon polarization
vector in the rest frame,
\be
S &=& (0, {\bf S}), \hspace{2em} {\bf S} \;\; = \;\; (0, 0, 1). 
\ee
\par
For spin--independent matrix elements such as $c^{(2)}$, 
eqs.(\ref{c2_def}), on the other hand, the flavor singlet part is 
leading in $N_c$. In the flavor singlet case $\hat{t} = 1$, the orientation 
matrices $R, R^\dagger$ are contracted, and the integral over rotations
becomes trivial,
\be
\int dR \, \phi_{S_3 T_3}^\dagger (R) \phi_{S_3 T_3} (R)  &=& 1 .
\label{int_rot_isosinglet}
\ee
One obtains
\be
\langle T_3 S_3 | 
\effop{O}^{S}_{\alpha_1 \ldots \alpha_r \beta_1 \ldots \beta_s} 
| T_3 S_3 \rangle_{\rm spin-dep.} &=& 2 M_N \; 
{\cal M}_{\alpha_1 \ldots \alpha_r \beta_1 \ldots \beta_s} ,
\nonumber 
\ee
\be
{\cal M}_{\alpha_1 \ldots \alpha_r \beta_1 \ldots \beta_s}
&=& (\ref{O_NS_levels}) \;\;\; \mbox{with}
\;\;\; \tau^a \rightarrow 1.
\label{O_S_levels}
\ee
\par
The isosinglet spin--dependent and isovector spin--independent
matrix elements are zero in the leading order of the $1/N_c$--expansion.
They appear only after including rotational corrections 
to eq.(\ref{O_general_levels}), {\em i.e.}, integrating over time--dependent 
rotations and expanding in the angular frequency, which is of order
$1/N_c$.
\par
The $N_c$--ordering of the matrix elements of twist--3 and 4 
operators is analogous to the one of the twist--2 quark distribution 
functions, where the isovector--polarized and isosinglet--unpolarized 
distributions appear in the leading order of the $1/N_c$--expansion.
The connection between spin and isospin is a general consequence 
of the hedgehog symmetry of the pion field, eq.(\ref{hedge}).
For reference, we have listed order of the various matrix elements 
in $N_c$ in table \ref{table_parametric}.
\par
We have obtained a closed expression for the matrix element of the
effective operator corresponding to a general gluonic operator
eq.(\ref{O}), in the form of a sum over quark single--particle
levels in the pion field of the soliton. This representation
is useful for studying the $N_c$--order of different
spin--isospin combinations. Moreover, eq.(\ref{O_general_levels})
could serve as a starting point for numerical calculations
of the matrix elements, using the quark wave functions determined 
by numerical diagonalization of the Dirac Hamiltonian, 
eq.(\ref{dirac}), in a finite box \cite{Review,DPPPW97}.
For further theoretical analysis, however, it is convenient
to pass to the more familiar language of Feynman diagrams.
Substituting in eq.(\ref{O_general_levels}) the spectral representation 
of the quark Green function in the background pion field,
\be
G(x , y ) &=& 
\theta (x_4 - y_4 ) \sum_{E_n > 0} \exp \left[ -E_n (x_4 - y_4 ) \right]
\Phi_n ({\bf x}) \Phi_n^\dagger ({\bf y}) i \gamma_4 \nonumber \\
&-&
\theta (y_4 - x_4 ) \sum_{E_n < 0} \exp \left[ -E_n (x_4 - y_4 ) \right]
\Phi_n ({\bf x}) \Phi_n^\dagger ({\bf y}) 
i \gamma_4 ,
\label{spectral}
\ee
and rearranging the terms in the sum over quark levels, one can
show that eq.(\ref{O_general_levels}) is equal to the sum of three 
terms which correspond to the Feynman diagrams for the 3--point 
correlation function shown in Fig.\ref{fig_nucleon}, 
\be
\lefteqn{ \langle T_3, S_3 | 
\effop{O}^{NS}_{\alpha_1 \ldots \alpha_r \beta_1 \ldots \beta_s} 
| T_3, S_3 \rangle_{\rm spin-dep.} } && \nonumber \\
&=& -2M_N (2 T_3) \frac{S_a}{3} 
\left[ {\cal M}_{(i)}^a + {\cal M}_{(ii)}^a + {\cal M}_{(iii)}^a  
\right]_{\alpha_1 \ldots \alpha_r \beta_1 \ldots \beta_s} .
\ee
We write them down for the spin--dependent 
isovector matrix element, eq.(\ref{O_NS_levels}); the corresponding 
expressions for the spin--independent isoscalar matrix elements are 
obtained by the trivial substitution eq.(\ref{O_S_levels}):
\par
{\em (i)} A ``valence quark'' contribution in which the effective operator 
connects to one valence quark line:
\be
\lefteqn{ 
{\cal M}^a_{\alpha_1 \ldots \alpha_r\beta_1 \ldots \beta_s , \, (i)} }
&& \nonumber \\
&=& - \frac{N_c}{2 T} \int d^4 x \int d^4 z 
\cF_{+ , \mu\nu\beta_1 \ldots \beta_s} (x - z ) 
\nonumber \\
&\times& \left\{ 
\exp \left[ E_{\rm lev} (x_4 - z_4 ) \right] 
\Phi_{\rm lev}^\dagger ({\bf x}) \left[ \gamma_4 
\Gamma_{\alpha_1 \ldots \alpha_r} \tau^a 
G(x , z) \gamma_4 \sigma_{\mu\nu} 
\frac{1 + \gamma_5}{2} M U_+ ({\bf z} ) \right]  
\Phi_{\rm lev} ({\bf z} ) \right. \nonumber \\ 
&& + \left. \exp \left[ - E_{\rm lev} (x_4 - z_4 ) \right] 
\Phi_{\rm lev}^\dagger ({\bf z}) \left[ \gamma_4 
\sigma_{\mu\nu} 
\frac{1 + \gamma_5}{2} M U_+ ({\bf z})  
\tau^a \Gamma_{\alpha_1 \ldots \alpha_r} G(z , x) \gamma_4 \right]
\Phi_{\rm lev} ({\bf x}) \right\}
\nonumber \\
&+& (+ \rightarrow - ) .
\label{me_i} 
\ee
This contribution has the form of a quark self-energy diagram evaluated 
between bound--state wave functions.  This contribution is similar to 
the Lamb shift in atomic physics.
\par
{\em (ii)} A ``sea quark'' contribution, in which the effective operator is
not connected to the valence quark lines but to a closed quark loop
in the background pion field. In momentum representation it 
can be written as 
\be
\lefteqn{
{\cal M}^a_{\alpha_1 \ldots \alpha_r \beta_1 \ldots \beta_s , \, (ii)} }
&& \nonumber \\
&=& - \frac{N_c}{2} 
\fourint{q} \fourint{k_1} \fourint{k_2} \fourint{l_1} \fourint{l_2}
\nonumber \\
&\times& (2\pi )^4
\delta^{(4)}(q - l_1 + l_2 + k_1 - k_2) 2\pi \delta (q_4 )
\nonumber \\
&\times& \cF_{+ , \mu\nu\beta_1 \ldots \beta_s} (k_2 - k_1 ) 
F(l_1 ) F(l_2 ) 
\tr \left[ G(l_2 , k_2 ) \tau^a \Gamma_{\alpha_1 \ldots \alpha_r} 
G(k_1 , l_1 ) 
\sigma_{\mu\nu} 
\frac{1 + \gamma_5}{2} M \tilde{U}_+ ({\bf q}) \right] 
\nonumber \\
&+& (+ \leftrightarrow -) ,
\label{me_ii}
\ee
where the quark Green function in the background pion field in momentum
representation is defined as
\be
G(x , z ) &=& \fourint{k} \fourint{l} e^{i (k\cdot x - l\cdot z)}
G(k, l),
\ee
and 
\be
{\tilde U}_\pm ({\bf q}) &=& \int d^3 x \; e^{-i {\bf q} \cdot {\bf x} }\;
\left[ U_\pm ({\bf x}) - 1 \right] .
\ee
is the Fourier transform of the hedgehog pion field, eq.(\ref{hedge}).
In eq.(\ref{me_ii}) we have reinstated the form factors $F(k)$ which
cut the quark loop momentum.
\par
{\em (iii)} An ``interaction'' contribution, in which the 
effective operator connects two different valence quark lines:
\be
\lefteqn{ 
{\cal M}^a_{\alpha_1 \ldots \alpha_r \beta_1 \ldots \beta_s , \, (iii)} }
&& \nonumber \\
&=&  \frac{N_c}{2 T} \int d^4 x \int d^4 z 
\cF_{+ , \mu\nu\beta_1 \ldots \beta_s} (x - z ) 
\nonumber \\
&\times& \left[ \Phi_{\rm lev}^\dagger ({\bf x}) \gamma_4 
\tau^a \Gamma_{\alpha_1 \ldots \alpha_r} \Phi_{\rm lev} ({\bf x})
\right] 
\left[ \Phi_{\rm lev}^\dagger ({\bf z}) \gamma_4 
\sigma_{\mu\nu} \frac{1 + \gamma_5}{2} M U_+ ({\bf z} )
\Phi_{\rm lev} ({\bf z}) 
\right] \nonumber \\ 
\nonumber \\
&+& (+ \rightarrow - ).
\label{me_iii}
\ee
This contribution describes interactions of pairs of valence
quarks in the nucleon mediated by the gluonic operator.
It is similar to the hyperfine splitting of baryon masses due 
to one--gluon exchange
\par
We emphasize that the expressions derived in this section apply to the 
leading order of the $1/N_c$--expansion. For matrix elements which are 
zero in leading order of the $1/N_c$--expansion one has to include 
rotational corrections.
\subsection{Higher--twist matrix elements in the naive quark model}
\label{subsec_naive}
Before evaluating the nucleon matrix elements of higher--twist operators
in the chiral soliton model it is useful to consider a much simpler 
picture of the nucleon, a naive constituent quark model. Assuming $N_c$
free constituent quarks in a spin--isospin state described by the 
non-relativistic quark model, one can easily construct the
nucleon matrix elements from the quark matrix elements derived in 
section \ref{subsec_quark_states}. We wish to caution the reader that,
as will be seen in the following section, the effects of the binding
of the quarks are essential, and the quark model results turn out
to be unrealistic; nevertheless, it is useful to have at hands these
simple estimates.
\par
In section \ref{subsec_quark_states} it was seen that the quark matrix
element $f^{(2)}_{\rm quark}$ can be reproduced by the local operator
eq.(\ref{f2_equiv}). For on-shell quarks the derivative in this operator
simply becomes $\partial^2 = M^2$, hence the operator reduces to the 
axial current. This immediately tells us that in the naive quark model
$f^{(2)}$ is proportional to the nucleon axial coupling 
constant (for both flavor--nonsinglet and singlet)
\be
f^{(2)}_{NS} &=& \frac{M^2}{M_N^2} \; f^{(2)}_{\rm quark} \; g_{A}^{(3)} ,
\hspace{2cm}
f^{(2)}_{S} \;\; = \;\; 
\frac{M^2}{M_N^2} \; f^{(2)}_{\rm quark} \; g_{A}^{(0)} .
\label{f2_quark_model}
\ee
The nucleon isovector axial coupling constant in the
large--$N_c$ quark model is $g_A^{(3)} = (N_c + 2 )/ 3 = O (N_c )$, 
while the isosinglet coupling constant $g_A^{(0)}$ is $O (N_c^0 )$, so 
the relative order in $1/N_c$ of $f^{(2)}_{NS}$ and $f^{(2)}_{S}$ 
is in agreement with the general results of the last section.
With $f^{(2)}_{\rm quark}$ given by eq.(\ref{f2_quark}), 
the phenomenological value $g_A^{(3)} = 1.25$, and a value of 
$g_A^{(0)} = 0.3$ obtained from an analysis of polarized nucleon 
structure functions of \cite{EllisK95}, this would mean
positive values for $f^{(2)}$: 
$f^{(2)}_{NS} = 0.75 M^2 / M_N^2 , \;
f^{(2)}_{S} = 0.18 M^2 / M_N^2$.
\par
Similarly, for the spin--dependent twist--3 matrix element $d^{(2)}$, 
eq.(\ref{d2_def}), one obtains in the naive quark model
\be
d^{(2)}_{NS} &=& \frac{M^2}{M_N^2} \; d^{(2)}_{\rm quark} \; g_{A}^{(3)} ,
\hspace{2cm}
d^{(2)}_{S} \;\; = \;\; \frac{M^2}{M_N^2} \; d^{(2)}_{\rm quark} \; 
g_{A}^{(0)} ,
\label{d2_quark_model}
\ee
which with eq.(\ref{d2_quark}) would imply $d^{(2)}_{NS} = -0.013 \;
M^2/M_N^2$ and $d^{(2)}_{S} = -3.3\times 10^{-3} M^2/M_N^2$.
\par 
The spin--independent matrix element $c^{(2)}$, eq.(\ref{c2_def}), 
is at quark level related to the matrix element of the vector 
current--type operator, eq.(\ref{c2_equiv}). Thus, in analogy to 
the spin--dependent case one finds that in the naive quark model 
$c^{(2)}_S$ is related to the nucleon vector charge (baryon number), 
which is $N_c$, while $c^{(2)}_{NS}$ is proportional to the nucleon 
isospin, which is of order $O(N_c^0 )$,
\be
c^{(2)}_{S} &=& \frac{M^2}{M_N^2} \; c^{(2)}_{\rm quark} \; N_c ,
\hspace{2cm}
c^{(2)}_{NS} \;\; = \;\; \frac{M^2}{M_N^2} \; c^{(2)}_{\rm quark} .
\label{c2_quark_model}
\ee
\par
The quark model estimates are useful since they illustrate the 
$N_c$--ordering of the different matrix elements and their parametric
magnitude in $M\rho$. We shall see, however, that the effects of the
binding of the quarks in the nucleon cannot be neglected because of 
the momentum dependence (non-locality) of the effective operators.
\subsection{The leading higher--twist matrix elements in the 
large--$N_c$ limit}
\label{subsec_nucleon_leading}
We now evaluate the leading twist--3 and 4 matrix elements in the 
large--$N_c$ limit in the chiral soliton model of the nucleon. As was 
shown in section \ref{subsec_nucleon_effective} the leading 
matrix elements are the flavor--nonsinglet spin--dependent matrix 
elements $d_{NS}^{(2)}$ and 
$f_{NS}^{(2)}$, eqs.(\ref{d2_def}, \ref{f2_def}), and the 
flavor--singlet spin--independent matrix element $c_S^{(2)}$, 
eq.(\ref{c2_def}).
\par
In addition to the $1/N_c$--classification we must keep in mind that
our theory is based on the smallness of the parameter $M\bar\rho$.
In section \ref{subsec_quark_states} we have seen that at quark level
the twist--4 matrix elements $f^{(2)}$ and $c^{(2)}$ are parametrically
of order unity, while $d^{(2)}$ is suppressed by a factor of 
$(M\bar\rho )^2$. We expect the parametric order of the nucleon matrix 
elements to be the same as that of the quark matrix elements; this is
indeed the case, as will become clear below. We shall therefore 
now consider separately the parametrically large (twist--4) and small
(twist--3) matrix elements. For quick reference we have listed the 
parametric order of the various matrix elements in 
table \ref{table_parametric}.
\par
\begin{figure}
\vspace{-1cm}
\epsfxsize=15cm
\epsfysize=12cm
\epsffile{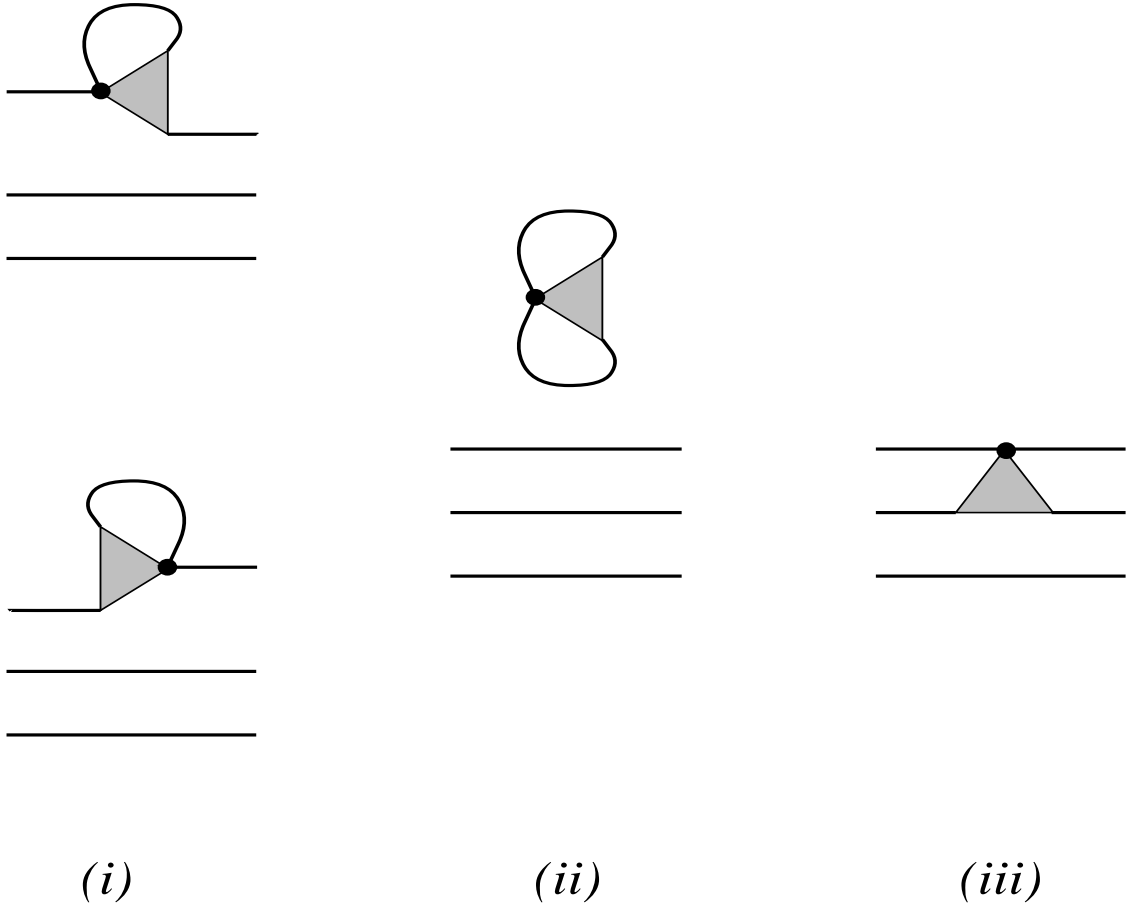}
\caption[]{Feynman diagrams contributing to the nucleon matrix element
of the effective operator. {\it (i)} ``Valence quark'' contribution, 
eq.(\ref{me_i}) {\it (ii)} ``Sea quark'' contribution, eq.(\ref{me_ii}). 
{\it (iii)} Interaction contribution, eq.(\ref{me_iii}).}
\label{fig_nucleon}
\end{figure}

Let us first consider the spin--dependent twist--4 matrix element
$f^{(2)}_{NS}$, eq.(\ref{f2_def}). A parametrically large contribution can 
come only from 
``divergent'' diagrams in Fig.\ref{fig_nucleon}, {\em i.e.}\ diagrams with 
characteristic momenta of the order of the cutoff, $\bar\rho^{-1}$. 
The diagrams containing divergences are the valence quark contribution
$(i)$ and the sea quark contribution $(ii)$, while
in the interaction diagrams $(iii)$ all momenta are cut 
by the bound--state level wave function. We shall therefore concentrate
on extracting the leading contribution in $M\bar\rho$ from diagrams
$(i)$ and $(ii)$.
\par
The valence quark contribution diagram $(i)$, eq.(\ref{me_i}), contains a 
quark ``self--energy'' diagram of the same type as appears in the 
on--shell quark matrix element, see Fig.\ref{fig_nucleon}.
In the nucleon case, of course, the momenta of the external quark 
lines are off--shell, as determined by the bound--state level wave function.
Since these momenta are of the order of the quark mass, $M$, it is sufficient
to compute the self--energy subdiagram for momenta of order $M$ ---
one does not need to know its full momentum dependence up to momenta
of order $\bar\rho^{-1}$. In other words, one may compute the 
contribution $(i)$ to the nucleon matrix element $f^{(2)}_{NS}$ 
using the local operators eq.(\ref{f2_equiv}); corrections to this 
local approximation are of
higher order in $M\bar\rho$. We thus estimate the contribution 
eq.(\ref{me_i}) to $f^{(2)}_{NS}$ as
\be
f^{(2)}_{NS, \, (i)} &=& \frac{M^2}{M_N^2} \; f^{(2)}_{\rm quark}
\; \widetilde{g}_{A, {\rm lev}}^{(3)} , 
\label{f2_ii}
\ee
where $\widetilde{g}_{A, {\rm lev}}^{(3)}$ denotes the matrix element of the 
isovector part of the operator eq.(\ref{f2_equiv}) between the 
bound--state wave functions\footnote{The explicit form of the bound--state
wave function can be found in ref.\cite{DPP88}.}, 
\be
\widetilde{g}_{A, {\rm lev}}^{(3)} &=& 
\frac{N_c}{9 M^2} \int d^3 x \; \Phi_{\rm lev}^\dagger ({\bf x}) \gamma_4 
\left[ \tau^a \gamma_a \gamma_5 
\left( E_{\rm lev}^2 + \vec\partial^2 \right) \right] 
\Phi_{\rm lev} ({\bf x}) , 
\label{ga_tilde} 
\ee
which is analogous to the level contribution to the axial coupling constant,
$g_{A, {\rm lev}}^{(3)}$, but contains the derivative operator. The effect 
of the latter are dramatic. Evaluating eq.(\ref{ga_tilde}) with the 
variational soliton profile eq.(\ref{varprof}) with $r_0 = 1.0/M$
we find $\widetilde{g}_{A, {\rm lev}}^{(3)} = -0.75$, while
the level contribution to the axial coupling constant is 
$g_{A, {\rm lev}}^{(3)} = 0.70$. Note that for a free quark, as was
assumed in the naive quark model estimate of section 
\ref{subsec_naive}, one would have 
$\widetilde{g}_{A, {\rm lev}}^{(3)} = g_{A, {\rm lev}}^{(3)}$. 
Thus, the binding of the quarks by the pion field reverses the sign
of the matrix element of the derivative operator eq.(\ref{f2_equiv}).
This fact is not dependent on the precise value of the
parameter $r_0$ which characterizes the radius of the classical pion 
field and thus the binding energy of the bound--state level; 
one finds that $\widetilde{g}_{A, {\rm lev}}^{(3)}$ becomes positive 
only for unphysical values $r_0 < 0.5/M$ where the level
is practically unbound. At the physical value $r_0 = 1.0/M$ we 
thus obtain, {\em cf.}\ eq.(\ref{f2_quark}),
\be
f^{(2)}_{NS, \, (i)} &=& -0.45 \times \frac{M^2}{M_N^2} .
\label{f2_i_num}
\ee
\par
In the sea quark contribution, diagram $(ii)$, eq.(\ref{me_ii}), on the 
other hand, all quark momenta are of the order of the cutoff, 
$\bar\rho^{-1}$, hence we must take into account the full momentum
dependence (non-locality) of the vertices in the effective operator.
To render an evaluation of this diagram feasible one can make use of 
an expansion of the quark propagators in the background pion field 
whose characteristic momentum is of order $M$. The expansion can 
be performed in such a way that it becomes 
exact in three limiting cases: {\em i}) small momenta, {\em ii}) 
large momenta, and {\em iii}) any momenta and small pion fields. This is 
the so--called interpolation formula of ref.\cite{DPPPW96}. Applying 
this expansion to eq.(\ref{me_ii}) we obtain
\be
f^{(2)}_{NS,\, (ii)} &=& 
\frac{M^2}{M_N^2} \frac{N_c}{2 \bar\rho^2} \;
\int \frac{d^3 q}{(2\pi )^3} I({\bf q}) \; q_a \;
\tr \left[ \tau^a \tilde{U}_- (-{\bf q}) \tilde{U}_+ ({\bf q}) \right] .
\label{f2_i}
\ee
where $I({\bf q})$ is given by a two--loop integrals with quark
propagators in the trivial background field ($U_\pm ({\bf x}) = 1$),
\be
I(q) &=& \bar\rho^4 
\fourint{k_1} \fourint{k_2} \;
\frac{{\cal G}(k) \, F(k_2 ) F (k_1) \, F^2 (k_1 + q)}
{\left( k_2^2 + M^2 F^4 \right) \left( k_1^2 + M^2 F^4 \right)
\left[ (k_1 + q)^2 + M^2 F^4 ) \right]} \nonumber \\
&& \times \left[ \frac{2}{3} N(k_1, k_2, q) + \frac{1}{3} 
N(k_1, k_2, 0) \right] ,
\ee
\be
N(k_1, k_2, q) &=& 
\frac{1}{3} \left( - k_1 \cdot (k_1 + q) + 
\frac{ 4 k\cdot k_1 \left[ k \cdot (k_1 + q) \right]}{k^2} \right) 
+ k\cdot (k_1 + q) , \nonumber \\ 
k &=& k_2 - k_1 .
\label{I_int}
\ee
Evaluating the integrals numerically at $M\bar\rho = 0.58$ we obtain 
\be
f^{(2)}_{NS,\, (ii)} &=& -0.27 \times \frac{M^2}{M_N^2} .
\label{f2_ii_num}
\ee
Adding the contributions $(i)$ and $(ii)$, eq.(\ref{f2_i_num}) and 
eq.(\ref{f2_ii_num}), using the standard value for the quark mass,
$M = 350\,{\rm MeV}$, and $M_N = 940\,{\rm MeV}$, we obtain our
estimate for the nucleon matrix element,
\be
f^{(2)}_{NS} &=& -0.72 \times \frac{M^2}{M_N^2}  \;\; = \;\; -0.10 .
\label{f2_final}
\ee
\par
The interaction contribution, diagrams $(iii)$, eq.(\ref{me_iii}),
is parametrically of order $(M\bar\rho )^2$. Explicit caclulation of 
the integral in eq.(\ref{me_iii}) shows that this contribution 
to $f^{(2)}_{NS}$ is numerically an order of magnitude smaller than 
the one from the ``divergent'' diagrams $(i)$ and $(ii)$ and can
safely be neglected. (We note, however, that in parametrically 
suppressed matrix elements such as $d^{(2)}$ the interaction 
contribution $(iii)$ is of the same order as $(i)$ and $(ii)$
and must be included.)
\par
It is interesting to consider the theoretical limit of large soliton
size, the ``skyrmion limit'', which may be regarded as the opposite 
of the naive quark model limit\footnote{For hadronic observables the 
two limits were investigated in ref.\cite{PBG95}.}. In this limit 
the valence quarks are strongly bound, {\em i.e.}, the energy of the 
discrete bound--state level approaches the lower continuum. In this case
diagrams $(ii)$ account for the entire contribution. Furthermore, one can
neglect the momentum dependence of the form factor $I(q)$ in 
eq.(\ref{f2_i}), in which case the function of the pion field in 
eq.(\ref{f2_i}) becomes proportional to the axial coupling 
constant, which for large solitons is given by
\be
g_A^{(3)} &=& -\frac{F_\pi^2}{9} \int d^3 x \; 
\tr \left[ -i \tau^a U^\dagger \partial_a U ({\bf x}) \right] ,
\ee 
where $F_\pi$ is the pion decay constant, which is defined by a
logarithmically divergent integral,
\be
F_\pi^2 &=& 4 N_c M^2 \fourint{k} \; 
\frac{F^4 (k)}{\left[ k^2 + M^2 F^4 (k) \right]^2} .
\label{Fpi}
\ee
(Numerically, $F_\pi = 93\, {\rm MeV}$ for $M\bar\rho = 0.58$.)
Thus, in the limit of large soliton size we obtain the relation
\be
f^{(2)}_{NS} &=& -\frac{M^2}{M_N^2} \;
\left( \frac{9 N_c \; I(0)}{2 \, \bar\rho^2 \, F_\pi^2} \right) \; 
g_A^{(3)} \;\; = \;\; -0.53 \times \frac{M^2}{M_N^2} \; g_A^{(3)} ,
\label{f2_NS_skyrmion}
\ee
similar to the naive quark model relation eq.(\ref{f2_quark_model}), 
but with a negative coefficient, since $I(0)$ is positive, 
$I(0) = 0.95\times 10^{-3}$. Quantitatively, with $g_A^{(3)} = 1.25$
this relation works well when compared to the result for physical soliton 
sizes, eq.(\ref{f2_final}).
\par
The calculation of the spin--independent twist--4 matrix element
$c^{(2)}_{S}$, eq.(\ref{c2_def}), proceeds in much the same way
as that of $f^{(2)}_{NS}$. This matrix element is also of order
unity in $M\bar\rho$, so the dominant contributions come from
diagrams $(i)$ and $(ii)$. The valence quark contribution
can again be estimated by ``localizing'' the self--energy 
subdiagram in $(i)$: 
\be
c^{(2)}_{S, \, (i)} &=& \frac{M^2}{M_N^2} \; c^{(2)}_{\rm quark}
\; \widetilde{g}_{V, {\rm lev}}^{(0)} , 
\label{c2_ii}
\ee
where $\widetilde{g}_{V, {\rm lev}}^{(0)}$ is the isosinglet matrix 
element of the local operator eq.(\ref{c2_equiv}) between
bound--state wave functions, 
\be
\widetilde{g}_{V, {\rm lev}}^{(0)} &=& 
\frac{N_c}{M^2} \int d^3 x \; \Phi_{\rm lev}^\dagger ({\bf x}) 
\left( E_{\rm lev}^2 + \vec\partial^2 \right)
\Phi_{\rm lev} ({\bf x}) . 
\label{gv_tilde} 
\ee
Without the derivatives this would be just the isosinglet vector
charge, {\em i.e.}, the baryon number, $N_c$. Numerically we find
$\widetilde{g}_{V, {\rm lev}}^{(0)} = -1.43 \times N_c$. 
The sea quark contribution $(ii)$ to $c^{(2)}_S$ emerges only in the 
third order of the expansion in pion field momenta (interpolation formula),
analogous to the expansion of the baryon charge, and is thus strongly
suppressed compared to the valence quark contribution for physical 
soliton sizes. Our result for the nucleon matrix element $c^{(2)}_S$ 
is therefore given entirely by eq.(\ref{c2_ii}),
\be
c^{(2)}_{S} &=& 2.6 \times \frac{M^2}{M_N^2} \;\; = \;\; 0.36 .
\label{c2_final}
\ee
Finally, we note that in the limit of large soliton size the nucleon
matrix element $c^{(2)}_{S}$ becomes proportional to the winding number
of the pion field,
\be
Q &=& - \frac{1}{24\pi^2} \int d^3 x \; \varepsilon_{ijk} \;
\tr \left[ (U^\dagger\partial_i U) (U^\dagger\partial_j U)
(U^\dagger\partial_k U)
\right] ,
\ee
which for large solitons coincides with the baryon number \cite{DPP88}.
One obtains
\be
c^{(2)}_S &=& \frac{M^2}{M_N^2} 
\left[ 72 \pi^2 \; (M\bar\rho )^2 \; I_{30} \right] N_c Q ,
\ee
where
\be
I_{30} &=& \fourint{k_1} \fourint{k_2} \;
\frac{{\cal G}(k) \; F^7 (k_2 ) F(k_1 )}
{\left[ k_2^2 + M^2 F^4 (k_2 ) \right]^4
 \left[ k_1^2 + M^2 F^4 (k_1 ) \right]}  N(k_1, k_2) ,
\nonumber 
\ee
\be
 N(k_1, k_2) &=& k_1^2
\left( - \frac{1}{3} + \frac{4}{3} \frac{(k\cdot k_1 )^2}{k^2 k_1^2} \right) 
+ k\cdot k_1 , 
\hspace{2em} k \;\; = \;\; k_2 - k_1
\label{I_30}
\ee
One finds that in the limit $M\bar\rho \rightarrow 0$
\be
(M\bar\rho )^2 I_{30} &\sim&  (M\bar\rho )^0 ,
\ee
in agreement with the fact that $c^{(2)}_S$ is parametrically of 
order $(M\bar\rho )^0$.
\par
Let us now turn to the spin--dependent twist--3 matrix element, 
$d^{(2)}_{NS}$, eq.(\ref{d2_def}). At quark level this matrix element
is proportional to $(M\bar\rho )^2$. At the present stage of
development of our theory we cannot make 
rigorous predictions at level $(M\bar\rho )^2$; this would require, 
among other things, to go beyond the one--instanton approximation in 
the effective operator. Nevertheless, it is worthwhile to make
a crude estimate of the nucleon matrix element by evaluating the 
one--instanton effective operator in the nucleon.
Since for parametrically suppressed matrix elements the effects of the
non-locality of the effective operator are important, we consider
the limit of large soliton size where the entire contribution
is given by diagrams $(ii)$. To compute $d^{(2)}_{NS}$ we 
contract eq.(\ref{me_ii}) with the 
``light--like'' Euclidean vector $n$ as in eq.(\ref{d2_projected}). 
This vector must now be taken in the nucleon rest frame, 
\be
n &=& (i, {\bf n})
\hspace{2em} {\bf n} \cdot {\bf S} \;\; = \;\; 0.
\ee 
We then proceed as in the case of $f^{(2)}_{NS}$, expanding the 
quark propagators in derivatives of the pion field. One finds that the 
first non-zero term appears at third order,
\be
d^{(2)}_{NS} &=& - \frac{M^2}{M_N^2} \frac{N_c}{6} 
\; (M\bar\rho )^2 \; I_{12} \; C[U] ,
\label{d2_NS_skyrmion} \\
C[U] &=& n_i n_j S_k S_a \nonumber \\
&& \times 
\int d^3 x \; \tr \left[ -i \tau^a (U^\dagger \partial_i U) 
\left[ (U^\dagger \partial_j U) (U^\dagger \partial_k U) 
- (U^\dagger \partial_j U) (U^\dagger \partial_k U) \right]
\; + \; (i\leftrightarrow j) \right] .
\nonumber
\\
\label{d2_structure}
\ee
Inserting the hedgehog pion field, eq.(\ref{hedge}), and averaging over 
orientations of the polarization vectors eq.(\ref{d2_structure})
becomes equal to
\be
C[U] &=& \frac{16 \pi^2}{3}
\int dr r^2 \; \left[ P^{'2} \frac{2 \sin P \cos P}{r} 
- P^{'} \frac{2 \sin^2 P}{r} + \frac{6 \sin^3 P \cos P}{r^3} 
\right] .
\ee
Numerically, $C[U] = 0.52 \times 10^2$ for the variational 
soliton profile of eq.(\ref{varprof}). In 
eq.(\ref{d2_NS_skyrmion}) $I_{12}$ is defined as
\be
I_{12} &=& \fourint{k_1} \fourint{k_2} \;
\frac{{\cal G}(k) \; F^3 (k_2 ) F(k_1 )^5}
{\left[ k_2^2 + M^2 F^4 (k_2 ) \right]^2
 \left[ k_1^2 + M^2 F^4 (k_1 ) \right]^3} N(k_1, k_2) ,
\ee
and one finds that
\be
(M\bar\rho )^2 I_{12} &\propto& M\bar\rho
\ee
for $M\bar\rho \rightarrow 0$, in agreement with the fact that 
at quark level $d^{(2)}$ is parametrically suppressed.
Numerically, at $M\bar\rho = 0.58$ one obtains 
$(M\bar\rho )^2 I_{12} = 0.43 \times 10^{-4}$, so that
\be
d^{(2)}_{NS} &=& -0.1 \times 10^{-2} \frac{M^2}{M_N^2} .
\ee
This number should be taken as an order--of--magnitude estimate.
\par
To summarize, we find that the nucleon matrix elements of higher--twist 
operators are of the same parametric order in $M\bar\rho$ as the 
corresponding matrix elements in quark states. (For reference, the 
parametric order of the different matrix elements is summarized in 
table \ref{table_parametric}.) Also, their numerical 
order of magnitude agrees with the quark model estimates. However,
we see that the binding of the quarks in the nucleon plays a decisive
role in the twist--4 nucleon matrix elements: the effective operators
at quark level are non-local (in the simplest form they may be 
approximated by local operators with derivative $\partial^2$), and the 
finite size of the bound state reverses the sign of the matrix element
as compared to free quarks. We conclude that for a reliable description
of twist--4 matrix elements it is of crucial importance that the
operators and the nucleon state are treated consistently within the
effective theory.
%
%
\noindent
\begin{table}[t]
\begin{center}
\[
\begin{array}{|l|c|c|c|}
\hline
\rule[-.75cm]{0cm}{1.5cm}
& \mbox{twist} & O(M\bar\rho ) & O(N_c ) \\ 
\hline
\hline
f^{(2)} & 4 & 1 & \begin{array}{rl} NS: & 1/N_c \\ S: & 1/N_c^2 \end{array} \\
\hline
d^{(2)} & 3 & (M\bar\rho )^2 &
     \begin{array}{rl} NS: & 1/N_c \\ S: & 1/N_c^2 \end{array} \\
\hline
\hline
c^{(2)} & 4 & 1 & \begin{array}{rl} S: & 1/N_c \\ NS: & 1/N_c^2 
\end{array} \\
\hline 
\end{array}
\]
\end{center}
\caption[]{A summary of the parametric order of the spin--dependent 
higher--twist matrix elements $f^{(2)}$ and $d^{(2)}$, eqs.(\ref{d2_def},
\ref{f2_def}), and the spin--independent matrix elements $c^{(2)}$,
eq.(\ref{c2_def}), in $1/N_c$ and the parameter $M\bar\rho$ related to the
packing fraction of the instanton medium, {\em cf.}\
eq.(\ref{M_rho_to_rho_R}). Details are given in the text.}
\label{table_parametric}
\end{table}
\section{Results and discussion}
\setcounter{equation}{0}
We now want to compare the numerical results for the twist--3 and 4
matrix elements with theoretical estimates obtained using other 
non-perturbative methods, as well as with estimates derived from
experimental data on polarized nucleon structure functions.
Numerical values are listed in table \ref{table_numeric}, where we 
also quote results of QCD sum rule calculations by 
Balitskii {\em et al.}\ \cite{BBK90} and Stein {\em et al}.\ 
\cite{Stein95t4,Stein95t3}, estimates obtained by Ji and Unrau using 
the bag model \cite{JiU94}, as well as lattice results from 
G\"ockeler {\em et al.}\ \cite{Goeckeler96}.
%
%
\noindent
\begin{table}[t]
\begin{center}
\[
\begin{array}{|l|c|c|c|c|c|c|c|}
\hline
& f^{(2)}_{NS} & d^{(2)}_{NS} & c^{(2)}_{S} & \mbox{scale}/{\rm GeV}^2 \\
\hline
\mbox{present work}            
& -0.10  & \sim 10^{-3} & 0.36 & \sim 0.4 \\
\mbox{Balitskii} \; \cite{BBK90}  
&  -0.20  &       0.072  & \mbox{---}  & 1   \\
\mbox{Stein} \; \cite{Stein95t4,Stein95t3}      
& -0.072 &        0.072  & \mbox{---}  & 1   \\
\mbox{Ji} \; \cite{JiU94}         
&  0.11  &        0.063  & \mbox{---}   & 5 \\
\mbox{G\"ockeler} \; \cite{Goeckeler96} 
& \mbox{---}  &  -0.13   & \mbox{---}   & 4 \\
\mbox{Braun} \; \cite{BK87}
& \mbox{---}  & \mbox{---} & 0.37 & 1 \\
\mbox{E142, E143, E154} \; \cite{Abe_p,Abe_n}                    
& \mbox{---}  &  0.043 \pm 0.046 & \mbox{---} & 3 \\
\mbox{Ji} \; \cite{JM97}         
& 0.10 \pm 0.28  & \mbox{---} & \mbox{---}  & 1 \\
\hline
\end{array}
\]
\end{center}
\caption[]{Summary of numerical results for the flavor--nonsinglet
spin--dependent twist--4 and 3 matrix elements $f^{(2)}_{NS}$ and 
$d^{(2)}_{NS}$, eqs.(\ref{d2_def}, \ref{f2_def}), and the 
flavor--singlet spin--independent twist--4 matrix element $c^{(2)}_{S}$, 
eq.(\ref{c2_def}). Flavor--nonsinglet and singlet parts are defined as in 
eqs.(\ref{a2_NS}, \ref{a2_S}). Shown are the results of the present 
calculation (for details see Sections \ref{subsec_nucleon_leading}), 
of QCD sum rule calculations by Balitskii {\em et al.}\ \cite{BBK90},
Stein {\em et al}.\ \cite{Stein95t4,Stein95t3} and by Braun and 
Kolesnichenko \cite{BK87}, results of a bag model calculation by 
Ji and Unrau \cite{JiU94}, as well as lattice calculations by 
G\"ockeler {\em et al.}\ \cite{Goeckeler96}.
Also shown are estimates of $d^{(2)}_{NS}$ based on 
measurements of the structure function $g_2$ by E142, E143, E154
\cite{Abe_p,Abe_n,Abe_measure} (for details see the text), and 
estimates of $f^{(2)}_{NS}$ from a recent analysis of power corrections
to $g_1$ by Ji and Melnichouk \cite{JM97}.}
\label{table_numeric}
\end{table}
\par
When discussing numerical values we must keep in mind that the 
theoretical status of our numbers presented in table \ref{table_numeric} 
depends on their parametric order, see table \ref{table_parametric}. 
The results for quantities of order unity 
in $M\bar\rho$, $f^{(2)}_{NS}$ and $c^{(2)}_{NS}$, may be regarded as 
quantitative estimates; the numerical value quoted for the parametrically 
suppressed twist--3 matrix element, $d^{(2)}_{NS}$, should be taken
as a crude estimate. We can say confidently only that numerically 
the twist--3 matrix elements are more than an order of magnitude smaller 
than the twist--4 ones.
\par
Our estimate for $f^{(2)}_{NS}$ argees well with the results of QCD 
sum rule calculations of \cite{BBK90} and \cite{Stein95t4,Stein95t3}. 
In particular, the large non-singlet value for $f^{(2)}$, which was first 
observed in the sum rule calculation of Balitskii {\em et al.}\ \cite{BBK90}, 
can naturally be explained by the fact that the non-singlet part is leading
in the $1/N_c$--expansion. Also, our result for the spin--independent
matrix element $c^{(2)}_{S}$ (corrections to the GLS sum rule) is in good 
agreement with the 
sum rule calculation of Braun and Kolesnichenko \cite{BK87}. In their
calculation the flavor--singlet part was found to be large, which, again, 
is consistent with the $1/N_c$--expansion.
We also note that our result for $f^{(2)}_{NS}$ differs in sign from the
bag model value of ref.\cite{JiU94}.
\par
As to the twist--3 matrix element, $d^{(2)}_{NS}$, we note that
our small value (``small'' on the typical scale of the twist--4 
matrix elements,
which is $M^2/M_N^2 \simeq 10^{-1}$) are consistent with the results of 
measurements of the polarized structure function $g_2$ 
\cite{Abe_p,Abe_n,Abe_measure}, where 
$d^{(2)}$ enters in the non-power suppressed--part, 
eqs.(\ref{g_2_NS}, \ref{g_2_S}). (The flavor--nonsinglet value 
we quote in table \ref{table_numeric} has been combined from 
the E143 value for $d^{(2)}_p$ \cite{Abe_p} and the SLAC average
for $d^{(2)}_n$ \cite{Abe_n} based on measurements 
of E142, E143 and E154 \cite{Abe_measure}, at a scale 
of $3 \, {\rm GeV}^2$.) In contrast, the lattice results 
of \cite{Goeckeler96} seem to overestimate these quantities.
\par
Finally, we note that using the numerical results of table 
\ref{table_numeric} one can establish the $Q^2$--dependence of the 
Bjorken, Ellis--Jaffe and Gross--Llewellyn-Smith sum rules. For this 
one has to add to
eqs.(\ref{g1_NS}, \ref{g1_S}) the logarithmic corrections, which can be
found in \cite{SV82}. Let us mention also that our value for 
$f^{(2)}_{NS}$ is within the bounds obtained from a phenomenological
analysis of power corrections to $g_1$ by Ji and Melnichouk \cite{JM97}.
\section{Conclusions and outlook}
\setcounter{equation}{0}
In this paper we have reported results of a comprehensive investigation 
of matrix elements of operators of twist 3 and 4 in the instanton vacuum. 
We have seen that, generally, the instanton vacuum provides a clear and 
consistent picture of the role of non-perturbative gluonic degrees of 
freedom in DIS matrix elements. The crucial element is the small parameter 
$M\bar\rho \propto (\bar\rho / R )^2$  inherent in this picture.
In particular, the fact that the QCD equations of motion are realized
at the level of effective operators provides a check for the consistency
of this approach.
\par
A marked distinction between operators of highest twist (twist 4, in our 
case) and lower twists has emerged. The contribution of the instanton
field in operators of twist 2 and 3 is suppressed
by a factor of the instanton packing fraction, $(M\bar\rho )^2$.
This justifies dropping the gluon field in twist--2 operators 
({\em i.e.}, replacing covariant by pure derivatives) when computing
twist--2 quark and antiquark distributions in the effective chiral 
theory \cite{DPPPW96,DPPPW97}. We see at present no 
other approach which provides an equally clear --- that is, parametric ---
justification for this ``quarks--antiquarks only'' approximation 
which proves to be very successful in practice.
On the other hand, instantons make an order unity contribution in 
operators of twist 4. The most striking example is the twist--4 operator 
of Section \ref{subsec_eom} which is zero by the QCD equations of motion: 
in the instanton vacuum it vanishes because of an $(M\bar\rho )^0$ 
instanton contribution.
\par
The deeper reason why instantons contribute only to the highest twist 
considered here is the $O(4)$ symmetry of the instanton, as a result of 
which the instanton can contribute only to the lowest--spin projection of a 
given tensor operator. To which twist this corresponds depends on the 
situation considered. Thus, rather than speak about twist one should 
say that instantons prefer lowest spin.
\par
As a result of the parametric suppression of twist 3 relative to twist 4
we obtain $d^{(2)} \ll f^{(2)}$ also numerically. (Generally, a positive 
feature of our approach is that the parametric ordering of quantities is 
borne out by the numerical values, {\em i.e.}, 
that parametrically suppressed contributions are also small numerically.)
The small value for $d^{(2)}_{NS}$ is consistent with present 
experimental data on polarized structure functions. 
\par
The numerical magnitude of the higher--twist matrix elements 
obtained in our approach can be understood in simple terms.
The constituent quark picture implied by the instanton vacuum means
that the typical magnitude of the nucleon matrix elements is 
$M^2/M_N^2$ times a number of order unity for parametrically
large matrix elements (twist 4), or a number $\ll 1$ for parametrically
suppressed matrix elements (twist 3). 
\par
We have found that the flavor--nonsinglet part of power corrections
to polarized structure functions, $f_{NS}^{(2)}$ and $d_{NS}^{(2)}$, 
is leading in the $1/N_c$--expansion. In the unpolarized case the situation 
is opposite, 
and the flavor--singlet part is leading. This ordering agrees well with 
the results of QCD sum rule calculations of \cite{BK87,BBK90}
\par
To summarize, the instanton vacuum implies a definite hierarchy of 
higher--twist matrix elements. With increasing accuracy of the 
measurements of polarized and unpolarized structure functions one should 
be able to test the predictions of this picture more accurately.
\par
The methods developed in this paper can be applied to calculate a 
wide range of other matrix elements relevant to DIS. For example, one can 
compute the twist--3 contribution to 
higher moments of $g_2$, eqs.(\ref{g_2_NS}, \ref{g_2_S}), and reconstruct
the entire structure function. Furthermore, the formalism developed in 
section \ref{subsec_nucleon_effective} can be applied to the 
calculation of nucleon matrix elements of 4--fermionic higher--twist
operators which appear in $1/Q^2$-- and $1/Q^4$--power corrections to 
unpolarized DIS \cite{SV82}.
\newpage
\noindent
{\large\bf Acknowledgements} \\[.3cm]
We are deeply grateful to D.I.\ Diakonov, V.Yu.\ Petrov and P.V.\ Pobylitsa 
for valuable suggestions and many helpful conversations. We wish to thank 
Klaus Goeke for encouragement and multiple help.
\\[.2cm]
This work has been supported in part by the Deutsche Forschungsgemeinschaft,
by a joint grant of the Deutsche Forschungsgemeinschaft and the Russian 
Foundation for Basic Research, and by COSY (J\"ulich). 
M.V.P.\ is supported by the A.v.Humboldt Foundation. 
%

%
%
\newpage
\newpage
\end{document}